\documentclass[12pt]{article}
\newlength{\dinwidth}
\newlength{\dinmargin}
\usepackage{amssymb}

\def\1#1{{\bf #1}}
\def\2#1{{\cal #1}}
\def\3#1{{\sl #1}}
\def\4#1{{\tt #1}}
\def\5#1{{\sf #1}}
\def\6#1{{\mathfrak #1}}
\def\7#1{{\mathbb #1}}

\newcommand{\be}{\begin{equation}}
\newcommand{\ee}{\end{equation}}
\newcommand{\ba}{\begin{array}}
\newcommand{\ea}{\end{array}}
\newcommand{\bea}{\begin{eqnarray}}
\newcommand{\eea}{\end{eqnarray}}
\newcommand{\bean}{\begin{eqnarray*}}
\newcommand{\eean}{\end{eqnarray*}}
\newcommand{\nn}{\nonumber}

\newcommand{\ve}{\varepsilon}
\newcommand{\vp}{\varphi}
\newcommand{\impl}{\Rightarrow}
\newcommand{\restr}{\upharpoonright}
\newcommand{\ol}{\overline}

\newcommand{\qed}{\ \ $\blacksquare$}
\newcommand{\qft}{quantum field theory}
\newcommand{\qfts}{quantum field theories}

\newcommand{\npb}{Nucl. Phys. \1B}
\newcommand{\cmp}{Commun. Math. Phys. }

\newcommand{\rmp}{Rev. Math. Phys. }
\newcommand{\jfa}{J. Funct. Anal. }

\newtheorem{defin}{Definition}[section]
\newtheorem{lemma}[defin]{Lemma} \newtheorem{prop}[defin]{Proposition}
\newtheorem{theorem}[defin]{Theorem}
\newtheorem{coro}[defin]{Corollary}

\newcommand{\bdefin}{\begin{defin}}
\newcommand{\blemma}{\begin{lemma}}
\newcommand{\bprop}{\begin{prop}}
\newcommand{\btheor}{\begin{theorem}}
\newcommand{\bcoro}{\begin{coro}}
\newcommand{\edefin}{\end{defin}}
\newcommand{\elemma}{\end{lemma}}
\newcommand{\eprop}{\end{prop}}
\newcommand{\etheor}{\end{theorem}}
\newcommand{\ecoro}{\end{coro}}

\newcommand{\poinc}{Poincar\'{e}}
\newcommand{\prf}{{\it Proof. }}
\newcommand{\rem}{{\it Remark. }}
\newcommand{\rems}{{\it Remarks. }}

\newcommand{\sectreset}[1]{\section{#1}\setcounter{equation}{0}
                             \message{\thesection : #1}}

\begin{document}\noindent
{\sf DESY 96-117} \hfill {\sf ISSN 0418-9833} \\
{\sf June 1996, revised May 1997} \hfill {\sf hep-th/9606175}

\begin{center} \vskip 14mm
{\LARGE Quantum Double Actions on Operator Algebras and}\\[3.2mm] 
{\LARGE Orbifold Quantum Field Theories}\\
\vskip 15mm
{\large Michael M\"uger\footnote{Supported by the Studienstiftung des deutschen Volkes}} \\[5mm]
{II.\ Institut f\"ur Theoretische Physik, Universit\"at Hamburg\\[1mm]
Luruper Chaussee 149, D--22761~Hamburg, Germany\\
Email: mueger@x4u2.desy.de \\[4mm]
June 26, 1996}
\end{center} 
\vskip 10mm

\abstract{Starting from a local quantum field theory with an unbroken compact symmetry 
group $G$ in $1+1$-dimensional spacetime we construct disorder fields implementing
gauge transformations on the fields (order variables) localized in a wedge region.
Enlarging the local algebras by these disorder fields we obtain a nonlocal field theory,
the fixpoint algebras of which under the appropriately extended action of the group $G$
are shown to satisfy Haag duality in every simple sector. 
The specifically 1+1 dimensional phenomenon of violation of Haag duality of
fixpoint nets is thereby clarified. In the case of a finite group $G$ the extended 
theory is acted upon in a completely canonical way by the quantum double $D(G)$ and
satisfies R-matrix commutation relations as well as a Verlinde algebra. Furthermore,
our methods are suitable for a concise and transparent approach to bosonization.
The main technical ingredient is a strengthened version of the split property which 
is expected to hold in all reasonable massive theories. In the appendices (part
of) the results are extended to arbitrary locally compact groups and our
methods are adapted to chiral theories on the circle.}

\sectreset{Introduction}
Since the notion of the `quantum double' was coined by Drinfel'd in his famous
ICM lecture \cite{drin1} there have been several attempts aimed at a clarification
of its relevance to two dimensional \qft. The quantum double appears implicitly
in the work \cite{dvvv} on orbifold constructions in conformal field theory, where
conformal \qfts\ (CQFTs) are considered whose operators are fixpoints under the 
action of a symmetry group on another CQFT. Whereas the
authors emphasize that `the fusion algebra of the holomorphic G-orbifold
theory naturally combines both the representation and class algebra of the group G'
the relevance of the double is fully recognized only in \cite{dpr}. There the
construction is also generalized by allowing for an arbitrary 3-cocycle in
$H^3(G, U(1))$ leading only, however, to a quasi quantum group in the sense of 
\cite{drin2}. The quantum double also appears in the context of integrable \qfts, e.g.\
\cite{bern}, as well as in certain lattice models (e.g.\ \cite{szlvec}).
Common to these works is the role of disorder operators or `twist fields' which are 
`local with respect to $\2A$ up to the action of an element $g\in G$' \cite{dvvv}.
Finally, it should be mentioned that the quantum double and its twisted generalization
also play a role in spontaneously broken gauge theories in $2+1$ dimensions (for a 
review and further references see \cite{b-wp}).

Regrettably most of these works (with the exception of \cite{szlvec}) are not very
precise in stating the premises and the results in mathematically unambiguous
terms. For example it is usually unclear whether the `twist fields' have to be 
constructed or are already present in some sense in the theory one starts with. 
As a means to improve on this state of affairs we propose to take seriously the 
generally accepted conviction that the physical content of a  \qft\ can be recovered 
by studying the inequivalent representations (superselection sectors) of the algebra 
$\2A$ of observables (which in the framework of conformal field theory is known as
the chiral algebra). This point of view, put forward as early as 
1964 \cite{haag1} but unfortunately widely ignored, has proved fruitful for the model 
independent study of (not necessarily conformally covariant) \qfts, for reviews see 
\cite{haag2,K}. Using the methods of algebraic \qft\ we
will exhibit the mechanisms which cause the quantum double to appear in {\it every}
\qft\ with group symmetry in $1+1$ dimensions fulfilling (besides the usual assumptions
like locality) only two technical assumptions (Haag duality and split property, see 
below) but independent of conformal covariance or exact integrability.

As in \cite{dhr1} we will consider a \qft\ to be specified by a net of von Neumann 
algebras, i.e.\ a map
\be \2O\mapsto\2F(\2O) ,\label{fvono}\ee
which assigns to any bounded region in $1+1$ dimensional Minkowski space a von 
Neumann algebra (i.e.\ an algebra of bounded operators closed under hermitian 
conjugation and weak limits) on the common Hilbert space $\2H$ such that
isotony holds:
\be \2O_1\subset\2O_2 \ \impl \ \2F(\2O_1)\subset\2F(\2O_2) .\ee
The quasilocal algebra $\2F$ defined by the union
\be \2F = \ol{\bigcup_{\2O\in\2K}\2F(\2O)}^{\|\cdot\|} \ee
over the set $\2K$ of all double cones (diamonds) is assumed to be irreducible, 
i.e.\ $\2F'=\7C\11$ \footnote{In general $\2M'=\{X\in\2B(\2H)|XY=YX\,\forall Y\in\2M\}$
denotes the algebra of all bounded operators commuting with all operators in $\2M$.}

The net is supposed to fulfill Bose-Fermi commutation relations, i.e.\ any local
operator decomposes into a bosonic and a fermionic part $F=F_+ + F_-$ such that
for spacelike separated $F$ and $G$ we have
\be [F_+,G_+]=[F_+,G_-]=[F_-,G_+]=\{F_-,G_-\}=0 .\label{commrel}\ee
The above decomposition is achieved by
\be F_\pm = \frac{1}{2}(F\pm\alpha_-(F)) ,\ee
where $\alpha_-(F)=VFV$ and $V=V^*=V^{-1}$ is the unitary operator which
acts trivially on the space of bosonic vectors and like $-\11$ on the fermionic 
ones. To formulate this locality requirement in a way more convenient for later 
purposes we introduce the twist operation $F^t=ZFZ^*$ where
\be Z=\frac{1+iV}{1+i}, \ \ \ (\Rightarrow Z^2=V)\ee
which leads to $ZF_+Z^*=F_+,\ ZF_-Z^*=iVF_-$ implying $[F,G^t]=0$. 
The (twisted) locality postulate (\ref{commrel}) can now be stated simply as
\be \2F(\2O)^t\subset\2F(\2O')'   .\label{commrel2}\ee

\poinc\ covariance is implemented by assuming the existence of a (strong\-ly
continuous) unitary representation on $\2H$ of the \poinc\ group $\2P$ such that
\be \alpha_{(\Lambda,a)}(\2F(\2O))=Ad\, U(\Lambda,a)(\2F(\2O))=
   \2F(\Lambda\2O+a) .\ee
The spectrum of the generators of the translations (momenta) is required to be 
contained in the closed forward lightcone and the existence of a unique vacuum vector
$\Omega$ invariant under $\2P$ is assumed. Covariance under the conformal 
group, however, is {\it not} required.

Our last postulate (for the moment) concerns the inner symmetries of the 
theory. There shall be a compact group $G$, represented in a strongly
continuous fashion by unitary operators on $\2H$ leaving invariant the vacuum
such that the automorphisms $\alpha_g(F)=Ad\, U(g)(F)$ of $\2B(\2H)$ 
respect the local structure:
\be \alpha_g(\2F(\2O))=\2F(\2O) .\ee
The action shall be faithful, i.e.\ $\alpha_g\ne\mbox{id}\ \forall g\in G$. This is
no real restriction, for the kernel of the homomorphism $G\to\mbox{Aut}(\2F)$ can 
be divided out.
(Compactness of $G$ need not be postulated, as it follows \cite[Theorem 3.1]{dl} from 
the split property which will be introduced later.)
In particular, there is an element $k\in Z(G)$ of order 2 in the center of the group 
$G$ such that $V=U(k)$. This implies that the observables which are now defined 
as the fixpoints under the action of $G$
\be \2A(\2O) = \2F(\2O)^G = \2F(\2O)\cap U(G)' \label{avono}\ee
fulfill locality in the conventional untwisted sense. In 1+1 dimensions the
representations of the \poinc\ group and of the inner symmetries do not necessarily 
commute. In the appendix of \cite{mue3} it is, however, proved that in theories 
satisfying the distal split property
the translations commute with the inner symmetries whereas the boosts act by
automorphisms on the group $G_{\mbox{max}}$ of all inner symmetries. As we will 
postulate a stronger version of the split property in the next section the cited 
result applies to the situation at hand. What we still have to assume is that the one 
parameter group of Lorentz boosts maps the group $G$ of inner symmetries, which in
general will be a subgroup of $G_{\mbox{max}}$, into itself and commutes with $V=U(k)$.
This assumption is indispensable for the covariance of the fixpoint net $\2A$ as
well as of another net to be constructed later.

This framework was the starting point for the investigations in \cite{dhr1}
where in particular properties of the observable net (\ref{avono}) and its
representations on the sectors in $\2H$, i.e.\ the G-invariant subspaces, were 
studied, implicitly assuming the spacetime to be of dimension $\ge 2+1$. 
While it is impossible to do any justice to the deep analysis which 
derives from this early work (e.g.\ \cite{dhr2,dhr3,dhr4,dr1,rob,dr2} 
and the books \cite{haag2,K})
we have to sketch some of the main ideas in order to prepare the ground for our own
work in the subsequent sections. One important notion examined in \cite{dhr1} was
that of {\it duality} designating a certain maximality property in the sense that 
the local algebras cannot be enlarged (on the same Hilbert space) without violating 
spacelike commutativity. The postulate of twisted duality for the fields consists 
in strengthening the twisted locality (\ref{commrel2}) to
\be \2F(\2O)^t = \2F(\2O')'   \label{twduality} ,\ee
which means that $\2F(\2O')$, the von Neumann algebra generated by all 
$\2F(\2O_1)$, $\2O'\supset\2O_1\in\2K$ contains all operators commuting
with $\2F(\2O)$ after twisting. From this it has been derived 
\cite[Theorem 4.1]{dhr1} that duality holds
for the observables when restricted to a simple sector
\be (\2A(\2O)\restr\2H_1)'  = \2A(\2O')\restr\2H_1  \label{duality0} .\ee
A sector $\2H_1$ is called simple if the group $G$ acts on it via multiplication with
a character
\be U(g)\restr\2H_1  = \chi(g)\cdot\11\restr\2H_1 .\ee
Clearly the vacuum sector is simple.
Furthermore it has been shown \cite[Theorem 6.1]{dhr1} that the irreducible 
representations of the observables on the charge sectors in $\2H$ are strongly 
locally equivalent in the sense that
for any representation $\pi(A)=A\restr\2H_\pi$ and any $\2O\in\2K$ there is a 
unitary operator $X_\2O: \2H_0\to\2H_\pi$ such that
\be X_\2O \, \pi_0(A)=\pi(A) \, X_\2O \ \ \forall A\in\2A(\2O') .
   \label{dhr0}\ee

The fundamental facts (\ref{duality0}) and (\ref{dhr0}), which have come to be called
Haag duality and the DHR criterion, respectively, were taken as starting points in
\cite{dhr3,dhr4} where a more ambitious approach to the theory of superselection
sectors was advocated and developed to a large extent. The basic idea was that the
physical content of any \qft\ should reside in the observables and their vacuum
representation. All other physically relevant representations as well as unobservable
charged fields interpolating between those and the vacuum sector should be 
constructed from the observable data. The vacuum representation and the other 
representations of interest were postulated to satisfy
\bea \pi_0(\2A(\2O)) &=& \pi_0(\2A(\2O'))', \label{duality}\\
   \pi\restr\2A(\2O') &\cong& \pi_0\restr\2A(\2O')\ \ \forall\2O\in\2K ,\label{dhr}\eea
respectively. It may be considered as one of the
triumphs of the algebraic approach that it has finally been possible to prove
\cite[and references given there]{dr2}
the existence of an essentially unique net of field algebras with a unique compact 
group $G$ of inner symmetries such that there is an isomorphism between the 
monoidal (strict, symmetric) category of the superselection sectors satisfying 
(\ref{dhr}) with the product structure established in \cite{dhr3} and the category of 
finite dimensional representations of $G$. Before turning now to the two 
dimensional situation we should remark that the duality property (\ref{twduality}) 
upon which the whole theory hinges has been proved to hold for free 
massive and massless fields (scalar \cite{araki} and Dirac \cite{dhr1}) in $\ge 1+1$
dimensions (apart from the massless scalar field in two dimensions) as well as for
several interacting theories ($P(\phi)_2, Y_2$). Furthermore, there is a remarkable
link \cite{rob} between Haag duality and spontaneous symmetry breakdown. For the rest of
this paragraph we assume that only a subgroup $G_0$ of $G$ is unbroken, i.e.\ unitarily
implemented on $\2H$. Then the net $\2B(\2O)=\2F(\2O)^{G_0}$ satisfies Haag duality in
restriction to $\2H_0=\2H^{G_0}$ whereas $\2A(\2O)=\2F(\2O)^G$, being a true subnet of
$\2B$, does not. Yet, defining the {\it dual net} (on $\2H_0$) by
\be \2A^d(\2O)=\2A(\2O')' ,\label{dualnet}\ee
the fixpoint net $\2A$ still satisfies {\it essential duality}
\be \2A^d(\2O)=\2A^{dd}(\2O) .\label{essduality}\ee
(Haag duality, by contrast, is simply $\2A(\2O)=\2A^d(\2O)$.) Furthermore, one finds
$\2A^d(\2O)=\2B(\2O)$. These matters have been developed  further in \cite{dr2, bdlr} in
the context of reconstruction of the fields from the observables. 

In $1+1$ dimensions a large part of the analysis sketched above breaks down due to the 
following topological peculiarities of $1+1$ dimensional Minkowski space. Firstly, there
is a \poinc\ invariant distinction between left and right, i.e.\ for a spacelike vector
$x$ the sign of $x^1$ is invariant under the unit component of $\2P$. This fact accounts
for the existence of {\it soliton sectors} which have been studied rigorously in the
frameworks of constructive and general \qft, see \cite{fro1} and \cite{fre1,fre2,schl},
respectively. We intend to make use of the latter in a sequel to this work. 

In the present paper we focus on the other well known feature of the topology of $1+1$ 
dimensional Minkowski space, viz.\ the fact that the spacelike complement of a bounded
connected region (in particular, a double cone) consists of two connected components.
The implications of this fact are twofold. On one hand, in the adaption of the DHR
analysis \cite{dhr3,dhr4} based on (\ref{duality},\ref{dhr}) to $1+1$ dimensions 
\cite{frs1,khr,frs2} the permutation group $\2S_\infty$ governing the statistics is 
replaced by the {\it braid group} $\2B_\infty$, as anticipated, e.g., in \cite{fro2}. 
It is still not known by which structure the compact group appearing
in the higher dimensional situation has to be replaced if a completely general 
solution to this question exists at all.

Besides the appearance of braid group statistics the disconnectedness of $\2O'$ 
manifests itself also if one starts from a field net $\2F$ with unbroken symmetry group
$G$. It was mentioned above that in $\ge 2+1$ dimensions the restriction of the fixpoint
net $\2A$ to the simple sectors in $\2H$ satisfies Haag duality provided the field net
$\2F$ fulfills (twisted) Haag duality. Since questions of Haag duality have been studied
only in the framework of the algebraic approach the third peculiarity of \qfts\ in $1+1$
dimensions (besides solitons and braid group statistics/quantum symmetry) is less widely
known. We refer to the fact that the step from (\ref{twduality}) to (\ref{duality}) may
fail in 1+1 dimensions. This means that one cannot conclude from twisted duality of the
fields that duality holds for the observables in simple sectors, which in fact is 
possible only in conformal theories. The origin of this phenomenon is easily understood.
Let $\2O\in\2K$ be a double cone. One can then construct gauge invariant
operators in $\2F(\2O')$ which are obviously contained in $\2A(\2O)'$
but not in $\2A(\2O')$. This is seen remarking that the latter algebra, belonging 
to a disconnected region, is defined to be generated by the observable algebras 
associated with the left and right spacelike complements of $\2O$, respectively.
This algebra does not contain gauge invariant operators constructed using fields
localized in both components.

We now come to the plan of this paper. Our aim will be to explore the relation 
between a \qft\ with symmetry group $G$ in 1+1 dimensions and the fixpoint theory. 
In addition to the general properties of such a theory stated above, twisted duality 
(\ref{twduality}) is assumed to hold for the large theory. As explained above, in this 
situation duality of the fixpoint theory fails even in the case of unbroken group 
symmetry. Yet there is a local extension which satisfies Haag duality and one would
like to obtain a complete understanding of this dual net. To this end we will need one 
additional postulate concerning the causal independence of one-sided infinite regions
(wedges) which are separated from each other by a finite spacelike distance. This 
property rules out conformal theories and singles out a (presumably large) class of 
well-behaved massive theories. 
In Section 2 we prove the existence of unitary disorder operators which implement a 
global symmetry transformation on one wedge and act trivially on the spacelike complement
of a slightly larger wedge. Using these operators we will in Section 3 consider a
non-local extension $\hat{\2F}(\2O)$ of the field net $\2F(\2O)$. The fixpoint net
$\hat{\2A}(\2O)$ of the enlarged net $\hat{\2F}(\2O)$ under the action of $G$ is shown 
to coincide with the dual net $\2A^d(\2O)$ (\ref{dualnet})
in restriction to the simple sectors. In conjunction with
several technical results on actions of $G$ this leads to an explicit characterization
of the dual net. In Section 4 we will show that there is an action of the quantum
double $D(G)$ on the extended net $\hat{\2F}$ and that the spacelike commutation
relations are governed by Drinfel'd's $R$-matrix. Since massive free scalar fields 
satisfy all assumptions we made on $\2F$ this construction 
provides the first mathematically rigorous construction of \qfts\ with $D(G)$-symmetry 
for any finite group $G$. The quantum double may be considered a `hidden symmetry' of 
the original theory since it is uncovered only upon extending the latter. The
$D(G)$-symmetry is spontaneously broken in that only the action of the subalgebra
$\7CG\subset D(G)$ is implemented in the Hilbert space $\2H$.
In analogy to Roberts' analysis this might be interpreted as the actual reason for the 
failure of Haag duality for the fixpoint net $\2A$. 
The aim of the final Section 5 is to show that the methods introduced in the 
preceding sections are well suited for a discussion of Jordan-Wigner transformations
and bosonization in the framework of algebraic \qft.

Three appendices are devoted in turn to a summary of the needed facts on quantum groups
and quantum doubles, a partial generalization of our results to infinite compact groups 
and an indication how an analysis similar to Sections 2 to 4 can be 
done for chiral conformal theories on the circle.

\sectreset{Disorder Variables and the Split Property}
\subsection{Preliminaries}
For any double cone $\2O\in\2K$ we designate the left and right spacelike 
complement by $W^\2O_{LL}$ and $W^\2O_{RR}$, respectively. Furthermore we write
$W^\2O_L$ and $W^\2O_R$ for ${W^\2O_{RR}}'$ and ${W^\2O_{LL}}'$. These
regions are wedge shaped, i.e.\ translates of the standard wedges
$W_L=\{x\in\7R^2 \mid x^1<-|x^0|\}$ and $W_R=\{x\in\7R^2 \mid x^1>|x^0|\}$.
We will not distinguish between open and closed regions, for definiteness one
may consider $\2O$ and all W-regions as open. With these definitions we have
$\2O=W^\2O_L \cap W^\2O_R$ and $\2O'=W^\2O_{LL} \cup W^\2O_{RR}$ which
graphically looks as follows.
\be\ba{c}\begin{picture}(300,100)(-150,-50)\thicklines
\put(20,0){\line(-1,1){40}}
\put(20,0){\line(-1,-1){40}}
\put(-20,0){\line(1,1){40}}
\put(-20,0){\line(1,-1){40}}
\put(20,0){\line(1,1){40}}
\put(20,0){\line(1,-1){40}}
\put(-20,0){\line(-1,1){40}}
\put(-20,0){\line(-1,-1){40}}
\put(-5,-5){$\2O$}
\put(-70,-5){$W^\2O_{LL}$}
\put(45,-5){$W^\2O_{RR}$}
\put(-30,15){$W^\2O_L$}
\put(10,15){$W^\2O_R$}
\end{picture} \ea\ee

Whereas, as we have shown in the introduction, Haag duality for double cones is 
violated in the fixpoint theory $\2A$, one obtains the following weaker form of 
duality.
\bprop The representation of the fixpoint net $\2A$ fulfills duality for
wedges
\be \2A(W)'=\2A(W') \ee
and essential duality (\ref{essduality}) in all simple sectors.
\label{prop1}\eprop
\prf The spacelike complement of a wedge region is itself a wedge, thus
connected, whereby the proof of \cite[Theorem 4.1]{dhr1} applies, yielding the
first statement. The second follows from wedge duality via
\be \2A^d(\2O)=\2A(\2O')'=(\2A(W^\2O_{LL})\vee\2A(W^\2O_{RR}))'
   =\2A(W^\2O_R)\wedge\2A(W^\2O_L)  ,\ee
as locality of the dual net is equivalent to essential duality of $\2A$. \qed 

We will now introduce the central notion for this paper. 
\bdefin A family of {\it disorder operators} consists, for any 
$\2O\in\2K$ and any $g\in G$, of two unitary operators $U^\2O_L(g)$ and
$U^\2O_R(g)$ verifying 
\be\ba{ccc} \mbox{Ad}U_L^\2O(g)\restr\2F(W^\2O_{LL}) =
   \mbox{Ad}U_R^\2O(g)\restr\2F(W^\2O_{RR}) &=& \alpha_g ,\\
   \mbox{Ad}U_L^\2O(g)\restr\2F(W^\2O_{RR}) =
   \mbox{Ad}U_R^\2O(g)\restr\2F(W^\2O_{LL}) &=& \mbox{id} .
\ea\ee\label{def1}\edefin
In words: the adjoint action of $U^{\2O}_{L/R}(g)$ on fields located in the
left and right spacelike complements of $\2O$, respectively, equals the global group 
action on one side and is trivial on the other. As a consequence of (twisted) wedge 
duality we have at once
\be U_L^\2O (g)\in \2F(W^\2O_L)^t , \ U_R^\2O (g)\in \2F(W^\2O_R)^t .\ee
On the other hand it is clear that disorder operators cannot be contained in the
local algebras $\2F(\2O), \2F(\2O)^t$ nor in the quasilocal algebra $\2F$,
for in this case locality would not allow their adjoint action to be as stated
on operators located arbitrarily far to the left or right. Heuristically, assuming
$U(g)$ arises from a conserved local current via $U(g)=e^{i\int j^0(t=0, x) dx}$,
one may think of $U_L^\2O(g)$ as given by
\be U_L^\2O(g)=e^{i\int_{-\infty}^{x_0} j^0(x) dx} ,\ee
where integration takes place over a spacelike curve from left spacelike infinity 
to a point $x_0$ in $\2O$. The need for a finitely extended interpolation 
region $\2O$ arises from the distributional character of the current which
necessitates a smooth cutoff. We refrain from discussing these matters further 
as they play no role in the sequel. In massive free field theories disorder operators
can be constructed rigorously (e.g.\ \cite{str,adler}) using the CCR/CAR structure and
the criteria due to Shale. 

\blemma Let $U^\2O_{L,1}(g),U^\2O_{L,2}(g)$ be disorder operators associated
with the same double cone and the same group element. Then 
$U^\2O_{L,1}(g)=F\,U^\2O_{L,2}(g)$ with $F\in \2F(\2O)^t$ unitary. An analogous statement
holds for the right handed disorder operators.
\label{uniq}\elemma
\prf Consider $F=U^\2O_{L,1}(g)\,U^{\2O*}_{L,2}(g)$. By construction 
$F\in\2F(W_L^\2O)^t$. On the other hand 
$Ad\, F\restr\2F(W_{LL}^\2O) = id$ holds as $U_{L,1}^\2O(g)$ and 
$U_{L,2}^\2O(g)$ implement the same automorphism of $\2F(W_{LL}^\2O)$.
By (twisted) wedge duality we have $F\in\2F(W_R^\2O)^t$ and (twisted) duality for 
double cones implies $F\in\2F(\2O)^t$. \qed\\
\rems 1. This result shows that disorder operators are unique up to unitary elements
of $\2F(\2O)^t$, the twisted algebra of the interpolation region. The obvious fact
that $U^\2O_L(g)\,U^\2O_L(h)$ and $U(g)\,U^\2O_L(h)\,U(g)^*$ are disorder operators
for the group elements $gh$ and $ghg^{-1}$, respectively, implies that a family of
disorder operators constitutes a projective representation of $G$ with the cocycle
taking values in $\2F(\2O)^t$.\\
2. Later on we will consider only bosonic disorder operators, which leads to the stronger
result $F\in\2F(\2O)_+$.

For the purposes of the present investigation the mere existence of disorder operators
is not enough, for we need them to obey certain further restrictions. Our first aim
will be to obtain such operators by a construction which is model independent
to the largest possible extent, making use only of properties valid in any reasonable
model. To this effect we reconsider an idea due to Doplicher \cite{dopl} and
developed further in, e.g., \cite{dl,bdl}.
It consists of using the split property \cite{buwi} to obtain, for any $g\in G$ and
any pair of double cones $\Lambda=(\2O_1, \2O_2)$ such that 
$\ol{\2O_1}\subset\2O_2$, an operator $U_\Lambda(g)\in\2F(\2O_2)$
such that
\be U_\Lambda(g) F U_\Lambda(g)^*=U(g) F U(g)^* \ \forall F\in\2F(\2O_1).\ee
In order to be able to do the same thing with wedges we introduce our last 
postulate.
\bdefin An inclusion $A\subset B$ of von Neumann algebras is split \cite{dl}, 
if there exists a type-I factor $N$ such that $A\subset N\subset B$. A net of field 
algebras satisfies the `split property for wedges' if the inclusions
$\2F(W^\2O_{LL})\subset\2F(W^\2O_L)$ and $\2F(W^\2O_{RR})\subset\2F(W^\2O_R)$ are 
split for every double cone $\2O$. (In our case, where wedge duality holds,
the split property for one of the above inclusions entails the same for the other
as is seen by passing to commutants and twisting.)
\edefin
This property is discussed to some length in \cite{mue5} and shown to be fulfilled
for the free massive scalar and Dirac fields. In \qfts\ where there are lots of
cyclic and separating vectors for the local algebras by the Reeh-Schlieder theorem, 
the split property is equivalent \cite{dl} to the existence, for any double 
cone $\2O$, of a unitary operator
$Y^\2O: \2H\to\2H\otimes\2H$ implementing an isomorphism between
$\2F(W^\2O_{LL})\vee \2F(W^\2O_{RR})^t$ and the tensor product
$\2F(W^\2O_{LL})\otimes \2F(W^\2O_{RR})^t$ (in the sense of von Neumann algebras)
\be Y^\2O \, F_1 F_2^t \, Y^{\2O *} = F_1 \otimes F_2^t \ \
  \forall F_1\in\2F(W^\2O_{LL}), F_2\in\2F(W^\2O_{RR})  .\ee
That one of the algebras $\2F(W^\2O_{LL})$ and $\2F(W^\2O_{RR})$, which are 
associated with spacelike separated regions, has to be twisted in order for an 
isomorphism as above to exist is clear as in general these algebras do not commute 
while the factors of a tensor product do commute. Analogously, 
there is a spatial isomorphism between $\2F(W^\2O_{LL})^t\vee \2F(W^\2O_{RR})$ and 
$\2F(W^\2O_{LL})^t\otimes \2F(W^\2O_{RR})$ implemented by $\tilde{Y}^\2O$.
We will stick to the use of $Y^\2O$ throughout.
In order not to obscure the basic simplicity of the argument we assume for a moment
that the theory $\2F$ is purely bosonic, i.e.\ fulfills locality and duality
without twisting. Using the isomorphism implemented by $Y^\2O$ we then have the
following correspondences:
\be\ba{ccccc}
   \2F(W^\2O_{LL}) & \cong & \2F(W^\2O_{LL}) & \otimes & \11 \\
   \2F(W^\2O_{RR}) & \cong & \11 & \otimes & \2F(W^\2O_{RR}) \\
   \2F(W^\2O_L) & \cong & \2B(\2H) & \otimes & \2F(W^\2O_L) \\
   \2F(W^\2O_R) & \cong & \2F(W^\2O_R) & \otimes & \2B(\2H) ,
\label{isom}\ea\ee
whereas Haag duality for double cones yields
\be \2F(\2O)=\2F(W^\2O_L)\wedge\2F(W^\2O_R) \cong
   \2F(W^\2O_R) \otimes \2F(W^\2O_L) .\label{isom2}\ee
(Taking the intersection separately for both factors of the tensor product is valid
in this situation as can easily be proved using the lattice property of von Neumann
algebras $\2M\wedge\2N=(\2M'\vee\2N')'$ and the commutation theorem for tensor
products $(\2M\otimes\2N)'=\2M'\otimes\2N'$.)
We thus see that in conjunction with the well known fact \cite{dri} that the 
algebras associated with wedge regions are factors of type $III_1$ the split property
for wedges implies that the algebras of double cones are type $III_1$ factors, too.

The following property of the maps $Y^\2O$ will be pivotal for the considerations 
below. Given any unitary $U$ implementing a local symmetry 
(i.e.\ $U\2F(\2O)U^*=\2F(\2O) \ \forall\2O$) and
leaving invariant the vacuum ($U\Omega=\Omega$) the following identity holds:
\be Y^\2O \, U = (U\otimes U) \, Y^\2O  .\label{coprod0}\ee
For the construction of $Y^\2O$ as well as for the proof of (\ref{coprod0}) we
refer to \cite{dl,bdl}, the difference that those authors work with double cones 
being unimportant.

\subsection{Construction of Disorder Operators}
The operators $Y^\2O$ will now be used to obtain disorder operators. To this
purpose we give the following
\bdefin For any double cone $\2O\in\2K$ and any $g\in G$ we set
\be\ba{ccc} U^\2O_L(g) & = & Y^{\2O*} \, (U(g)\otimes \11) \, Y^\2O \\
   U^\2O_R(g) & = & Y^{\2O*} \, (\11\otimes U(g)) \, Y^\2O .
\ea\ee\label{def2}\edefin
As an immediate consequence of this definition we have the following
\bprop The disorder operators defined above satisfy
\bea \left[  U^{\2O}_L (g), U^{\2O}_R (h) \right]  &=&  0 ,\\
   U^{\2O}_L (g) \ U^{\2O}_R (g) & = & U(g) ,\label{factoriz}\eea
\be U(g)\, U^{\2O}_{L/R} (h)\, U(g)^* = U^{\2O}_{L/R} (ghg^{-1}) .\label{cov}\ee
\eprop
\prf The first statement is trivial and the second follows from (\ref{coprod0}).
The covariance property (\ref{cov}) is another consequence of (\ref{coprod0}). \qed\\
\rem We have thus obtained some kind of factorization of the global action of the 
group $G$ into two commuting {\em true} (i.e.\ no cocycles) representations of $G$ 
such that the original action is recovered as the diagonal. Furthermore, 
these operators transform covariantly under global gauge transformations. 
In particular they are bosonic since $k\in Z(G)$.

It remains to be shown that the $U^\2O_{L/R}$ indeed fulfill the requirements
of Definition \ref{def1}. The second requirement follows from Definition \ref{def2},
which with (\ref{isom}) obviously yields
\be U_L^\2O (g)\in \2F(W^\2O_L) , \ U_R^\2O (g)\in \2F(W^\2O_R) .\ee
The first one is seen by the following computation valid for 
$F\in\2F(W^\2O_{LL})$
\bea \lefteqn{U^\2O_L(g) F U^{\2O*}_L(g) \cong (U(g)\otimes\11)
  (F\otimes\11)(U(g)\otimes\11)^*} \\ 
 && = (U(g)FU(g)^*\otimes\11) \cong U(g)FU(g)^* \nn\eea
appealing to the isomorphism $\cong$ implemented by $Y^\2O$.

Returning now to the more general case including fermions we have to consider
the apparent problem that there are now two ways to define the operators 
$U^\2O_L(g)$ and $U^\2O_R(g)$, depending upon whether we choose $Y^\2O$ or 
$\tilde{Y}^\2O$. (By contrast, the tensor product factorization (\ref{isom2}) of 
the local algebras is of a purely technical nature, rendering it irrelevant whether we 
use $Y^\2O$ or $\tilde{Y}^\2O$.) 
This ambiguity is resolved by remarking that the element $k\in G$ giving rise to 
$V$ by $V=U(k)$ is central, implying that the operators $U(g),\ g\in G$, are bosonic 
(even). For even operators $F_1\in\2F(W^\2O_{LL}), F_2\in\2F(W^\2O_{RR})$ we have
$F_1=F_1^t, F_2=F_2^t$ and thus
\be Y^\2O\, F_1 F_2 \, Y^{\2O*} = \tilde{Y}^\2O\, F_1 F_2 \, \tilde{Y}^{\2O*} = 
   F_1\otimes F_2 \ee
so that the disorder variables are uniquely defined even operators. 

The first two equations of (\ref{isom}) are replaced by
\be\ba{ccccc}
   \2F(W^\2O_{LL}) & \cong & \2F(W^\2O_{LL}) & \otimes & \11, \\
   \2F(W^\2O_{RR})^t & \cong & \11 & \otimes & \2F(W^\2O_{RR})^t .
\label{newisom1}\ea\ee
By taking commutants we obtain
\be\ba{ccccc}
   \2F(W^\2O_L) & \cong & \2B(\2H) & \otimes & \2F(W^\2O_L), \\
   \2F(W^\2O_R)^t & \cong & \2F(W^\2O_R)^t & \otimes & \2B(\2H) 
\label{newisom2}\ea\ee
and an application of the twist operation to the second equations of (\ref{newisom1})
and (\ref{newisom2}) yields
\be\ba{ccccccccc} \2F(W^\2O_{RR})&\cong&\11&\otimes&\2F(W^\2O_{RR})_+ &+& 
   V&\otimes&\2F(W^\2O_{RR})_- ,\\
  \2F(W^\2O_R)&\cong& \2F(W^\2O_R)& \otimes& \2B(\2H)_+ &+& \2F(W^\2O_R)\,V &\otimes&
   \2B(\2H)_- .\label{newisom4}\ea\ee
The identity $\2F(\2O)=\2F(W^\2O_L)\wedge\2F(W^\2O_R)$, which is valid in the 
fermionic case, too, finally leads to
\be \2F(\2O)\cong \2F(W^\2O_R) \otimes \2F(W^\2O_L)_+ \ + \
   \2F(W^\2O_R)\,V \otimes \2F(W^\2O_L)_- . \label{newisom3}\ee
While this is not as nice as (\ref{isom2}) it is still sufficient for the 
considerations in the sequel. That $\2F(\2O),\ \2O\in\2K$ is a factor is, however, 
less obvious than in the pure Bose case and will be proved only in Subsection 
\ref{o-ahhat}.

The following easy result will be of considerable importance later on.
\blemma The disorder operators $U^\2O_L(g)$ and $U^\2O_R(g)$ associated with the double 
cone $\2O$ implement automorphisms of the local algebra $\2F(\2O)$. \elemma
\prf In the pure Bose case this is obvious from Definition \ref{def2}, (\ref{isom2}) 
and the fact that $Ad\,U(g)$ acts as an automorphism on all wedge algebras.
In the Bose-Fermi case (\ref{newisom3}) the same is true since $U(g)$ commutes
with $V=U(k)$. \qed
\bdefin $\alpha_g^\2O= Ad\,U^\2O_L(g),\ \ g\in G, \2O\in\2K$. \edefin

We close this section with one remark. We have seen that the split property for
wedges implies the existence of disorder operators which constitute true 
representations of the symmetry group and which transform covariantly under the global
symmetry. Conversely, one can show that the existence of disorder operators, possibly
with group cocycle, in conjunction with the split property for wedges for the 
fixpoint net $\2A$ implies the split property for the field net $\2F$. This in turn
allows to remove the cocyle using the above construction. We refrain from giving the
argument which is similar to those in \cite[pp.\ 79, 85]{dopl}.

\sectreset{Field Extensions and Haag Duality}
\subsection{The Extended Field Net}
Having defined the disorder variables we now take the next step, which at first sight 
may seem unmotivated. Its relevance will become clear in the sequel. 
We define a new net of algebras $\2O\mapsto\hat{\2F}(\2O)$
by adding the disorder variables associated with the double cone $\2O$ to
the fields localized in this region.
\bdefin\be \hat\2F(\2O)=\2F(\2O) \vee U^{\2O}_L (G)'' .\ee
\label{def3}\edefin
\rems 1. In accordance with the common terminology in statistical mechanics and
conformal field theory the operators which are composed of fields (order variables) 
and disorder variables might be called {\it parafermion operators}.\\
2. We could as well have chosen the disorder operators acting on the right hand side.
As there is a complete symmetry between left and right there would be no fundamental 
difference. We will therefore stick to the above choice throughout this paper. 
Including both the left and right handed disorder operators would, however, have the 
unpleasant consequence that there would be translation invariant operators (namely the
$U(g)$'s) in the local algebras. \\
3. The local algebra $\hat{\2F}(\2O)$ of the above definition resembles the crossed 
product of $\2F(\2O)$ by the automorphism group $\alpha_g^\2O$, 
the interesting aspect being that the automorphism group depends on the region $\2O$. 
These two constructions differ, however, with respect to the Hilbert space on
which they are defined. Whereas the crossed product 
$\2F(\2O)\rtimes_{\alpha^\2O} G$ lives on the Hilbert space $L^2(G,\2H)$, 
our algebras $\hat{\2F}(\2O)$ are defined on the original space $\2H$. 
For later purposes it will be necessary to know whether these algebras are 
isomorphic, but we prefer first to discuss those aspects which are independent
of this question.

The first thing to check is, of course, that the Definition \ref{def3} specifies a 
net of von Neumann  algebras.
\bprop The assignment $\2O\mapsto\hat{\2F}(\2O)$ satisfies isotony.
\label{isot}\eprop
\prf Let $\2O\subset\hat{\2O}$ be an inclusion of double cones. Obviously we have
$\2F(\2O)\subset\hat{\2F}(\hat{\2O})$. In order to prove
$U^\2O_L(g)\in\hat{\2F}(\hat{\2O})$ we observe that $U^\2O_L(g)$ is a disorder
operator for the larger region $\hat{\2O}$, too. Thus, by Lemma \ref{uniq}
we have $U^\2O_L(g)=F\,U^{\hat{\2O}}_L(g)$ with $F\in\2F(\hat{\2O})$.
Now it is clear that $U_L^\2O(g)\in\hat{\2F}(\hat{\2O})$. \qed\\
\rem From this we can conclude that the net $\hat{\2F}(\2O)$ is
uniquely defined in the sense that any family of bosonic disorder operators gives rise 
to the same net $\hat{\2F}(\2O)$ provided such operators exist at all. For most 
of the arguments in this paper we will, however, need the detailed properties proved
above which follow from the construction via the split property. 

It is obvious that the net $\hat{\2F}$ is nonlocal. While the spacelike commutation
relations of fields and disorder operators are known by construction we will have
more to say on this subject later. On the other hand it should be clear that the
nets $\hat{\2F}$ and $\2A$ are local relative to each other. This is simply the 
fact that the disorder operators commute with the fixpoints of $\alpha_g$ in both 
spacelike complements. 

\bprop The net $\hat{\2F}$ is \poinc\ covariant with the original
representation of $\2P$. In particular $\alpha_a(U^\2O_L(g))=U^{\2O+a}_L(g)$ 
whereas for the boosts we have
\be \alpha_\Lambda (U^\2O_L(g))=U^{\Lambda\2O}_L(h) ,\ee
if $U(\Lambda)\,U(g)\,U(\Lambda)^*=U(h)$ .\eprop
\prf 
The family $Y^\2O: \, \2H\to\2H\otimes\2H$ of unitaries provided by the
split property fulfills the identity
\be Y^{\Lambda\2O+a}=(U(\Lambda,a)\otimes U(\Lambda,a)) \,
   Y^\2O \, U(\Lambda,a)^* ,\ee
as is easily seen to follow from the construction in \cite{dl,bdl}. This implies
\bea \alpha_{\Lambda,a}(U^\2O_L(g))&=&U(\Lambda,a) \, {Y^\2O}^* \,
   (U(g)\otimes\11)\, Y^\2O \, U(\Lambda,a)^* \nn\\
   &=& {Y^{\Lambda\2O+a}}^*\,(U(\Lambda,a)\,U(g)\,U(\Lambda,a)^*\otimes\11)\,
   Y^{\Lambda\2O+a} \\
   &=& U_L^{\Lambda\2O+a} (h) ,\nn\eea
where $U(\Lambda)\,U(g)\,U(\Lambda)^*=U(h)$. \qed
\bprop The vacuum vector $\Omega$ is cyclic and separating for 
$\hat{\2F}(\2O)$.
\eprop
\prf Follows from 
\be \2F(\2O)\subset\hat{\2F}(\2O)\subset\2F(W^\2O_L) \ee
since $\Omega$ is cyclic and separating for $\2F(\2O)$ and $\2F(W^\2O_L)$.\qed
\bprop The wedge algebras for the net $\hat{\2F}$ take the form
\be \hat{\2F}(W^\2O_L)=\2F(W^\2O_L), \ \ 
   \hat{\2F}(W^\2O_R)=\2F(W^\2O_R)\vee U(G)''=\2A(W^\2O_{LL})' . \ee
As a consequence $\Omega$ is not separating for $\hat{\2F}(W^\2O_R)$!
\label{prop6}\eprop
\prf The first identity is obvious, while the second follows from 
$\2F(W^\2O_R)\ni U^{\hat{\2O}}_R(g)\ \forall\hat{\2O}\in W^\2O_R$
and the factorization property (\ref{factoriz}).
The last statement is equivalent to $\Omega$ not being cyclic
for $\2A(W^\2O_{LL})$. \qed
\bprop Let $\hat{F}\in\2F(\2O)U^\2O_L(g)$. Then the following cluster 
properties hold.
\bea w-\lim_{x\rightarrow -\infty} \alpha_x(\hat{F}) &=& 
  \langle\Omega,\hat{F}\Omega\rangle\cdot\11,\\
  w-\lim_{x\rightarrow +\infty} \alpha_x(\hat{F}) &=& 
   \langle\Omega,\hat{F}\Omega\rangle\cdot U(g) .\eea
\eprop
\prf The first identity follows from $\hat{F}\in\hat{\2F}(W^\2O_L)$ and the usual
cluster property. The second is seen by writing $\hat{F}=F\,U^\2O_R(g^{-1})\,U(g)$
and applying the weak convergence of $U^\2O_R$ as above, the translation invariance
of $U(g)$ and the invariance of the vacuum under $U(g)$.\qed

\subsection{Haag Duality}
Observing by (\ref{cov}) that the adjoint action of the global symmetry group 
leaves the `localization' (in the sense of Definition \ref{def1}) of the disorder 
operators invariant it is clear that the automorphisms $\alpha_g=Ad\, U(g)$ extend to 
local symmetries of the enlarged net $\hat{\2F}$. We are thus in a position 
to define yet another net, the fixpoint net of $\hat{\2F}$
\bdefin \be \hat\2A(\2O) = \hat{\2F}(\2O) \wedge U(G)' ,\ee
\label{def4}\edefin
whereby we have the following square of local inclusions
\be\ba{ccc} \hat{\2A}(\2O) & \subset & \hat{\2F}(\2O) \\ \cup & & \cup \\
     \2A(\2O) & \subset & \2F(\2O) 
.\ea\label{square}\ee
\rem The conditional expectation $m(\cdot)=\int dg\,\alpha_g(\cdot)$ from 
$\hat{\2F}(\2O)$ to $\hat{\2A}(\2O)$ clearly restricts to a conditional expectation from
$\2F(\2O)$ to $\2A(\2O)$. In Section 4 we will see that there is also a conditional
expectation $\gamma_e$ from $\hat{\2F}(\2O)$ to $\2F(\2O)$ which restricts to a
conditional expectation from $\hat{\2A}(\2O)$ to $\2A(\2O)$, provided the group $G$ is
finite. Since $\gamma_e$ commutes with $m$ the square (\ref{square}) then
constitutes a commuting square in the sense of Popa.

\bprop The net $\2O\mapsto\hat\2A(\2O)$ is local. \eprop
\prf Let $\2O < \tilde{\2O}$ be two regions spacelike to each other, 
$\tilde{\2O}$ being located to the right of $\2O$. From 
$\hat{\2A}(\2O)\subset\2A(W^\2O_L)$ and the relative locality of observables
and fields we conclude that $\hat{\2A}(\2O)$ commutes with $\2F(\tilde{\2O})$.
On the other hand the operators $U_L^{\tilde{\2O}}(g)$ commute with 
$\hat{\2A}(\2O)\subset\hat{\2F}(W^\2O_L)=\2F(W^\2O_L)$ since 
$Ad\,U_L^{\tilde{\2O}}(g)\restr\2F(W^\2O_L)=\alpha_g$ and $\hat{\2A}(\2O)$ is pointwise
gauge invariant. \qed

We have just proved that the net $\hat{\2A}$ constitutes a local extension of
the observable net $\2A$, thereby confirming our initial observation that
$\2A$ does not satisfy Haag duality. The elements of $\hat{\2A}$ being gauge
invariant they commute a fortiori with the central projections in the group algebra,
thereby leaving invariant the sectors in $\2H$. We will now prove a nice result 
which serves as our first justification for the Definitions \ref{def3} and \ref{def4}.
\blemma
\be \2A(\2O')' = \2F(\2O)\vee U^\2O_L(G)''\vee U^\2O_R(G)''.\ee
\label{lem1}\elemma
\prf We already know that
\be \2A(\2O')'\supset \2F(\2O)\vee U^\2O_L(G)''\vee U^\2O_R(G)''.\ee
In order to prove equality we consider the following string of identities, making
use of the spatial isomorphisms due to the split property and omitting the
superscript $\2O$ on the wedge regions.
\bea \2A(\2O')'& =&(\2A(W_{LL})\vee\2A(W_{RR}))' \nn\\
   &=&((\2F(W_{LL})\wedge U(G)')\vee(\2F(W_{RR})\wedge U(G)'))'\nn\\
   &\cong& \left((\2F(W_{LL})\wedge U(G)')\otimes(\2F(W_{RR})\wedge U(G)')
         \right)' \nn\\
   &=&(\2F(W_R)^t\vee U(G)'')\otimes(\2F(W_L)^t\vee U(G)'') ,\nn\\
   &=& (\2F(W_R)\otimes\2F(W_L))\vee(U(G)''\otimes\11)
   \vee(\11\otimes U(G)'') \nn\\
   &\cong& \2F(\2O)\vee U^\2O_L(G)''\vee U^\2O_L(G)'' . \eea
In the third step we have used the identities 
$\2F(W_{LL})\wedge U(G)'\cong\2F(W_{LL})\wedge U(G)'\otimes\11$ and
$\2F(W_{RR})\wedge U(G)'\cong\11\otimes\2F(W_{RR})\wedge U(G)'$ which are easily
seen to follow from (\ref{newisom2}) and (\ref{newisom4}), respectively. The fourth 
step is justified by $\2F(\cdot)^t\vee U(G)''=\2F(\cdot)\vee U(G)''$. \qed
\btheor In restriction to a simple sector $\2H_1$ the net
$\hat{\2A}$ satisfies Haag duality, i.e.\ it coincides with the dual net $\2A^d$ in
this representation. \label{thm1}\etheor 
\prf Let $P_1$ be the projection on a simple sector, i.e.\ fulfilling
\be U(g) P_1 = P_1 U(g) = \chi(g)\cdot P_1 ,\label{1dim}\ee
where $\chi$ is a character of $G$. Making use of
$U^\2O_L(G)''\vee U^\2O_R(G)''=U^\2O_L(G)''\vee U(G)''$ and (\ref{1dim}) we have
\bea \lefteqn{P_1 \, (\2F(\2O)\vee U^\2O_L(G)''\vee U^\2O_R(G)'') \, P_1=}\nn\\
   & & P_1 \, (\2F(\2O)\vee U^\2O_L(G)''\vee U(G)'') \, P_1 =
   P_1 \, \hat{\2F}(\2O) \, P_1  .\eea
With $m(F)=\int dg \, U(g) F U(g)^*$ and once again using (\ref{1dim}) we obtain
\be P_1 \, \hat{\2F}(\2O) \, P_1 = P_1 \, m(\hat{\2F}(\2O)) \, P_1 =  P_1 \, 
   \hat{\2A}(\2O) \, P_1 .\ee
On the other hand
\be P_1 \, \2A(\2O')' \, P_1\restr P_1 \2H =
  \left(P_1 \, \2A(\2O') \, P_1\restr P_1 \2H \right) '  .\ee
The proof is now completed by applying the preceding lemma. \qed\\
\rem The above arguments make it clear that Haag duality cannot hold for the net
$\2A(\2O)$ even in simple sectors. This is not necessarily so if the split
property for wedges does not hold. In conformally invariant theories gauge invariant 
combinations of field operators in the left and the right spacelike complements of 
a double cone $\2O$ may well be contained in $\2A(\2O')$ due to spacetime
compactification. One would think, however, that this is impossible in massive theories,
even those without the split property.

\subsection{Outerness Properties and Computation of $\hat{\2A}(\2O)$}\label{o-ahhat}
While this theorem allows us in principle to construct the dual net $\hat{\2A}$ one 
would like to know more explicitly how the elements of $\hat{\2A}$ look in terms 
of the fields in $\2F$ and the disorder operators. In the case of an abelian
group $G$ this is easy to see. As a consequence of the covariance property 
(\ref{cov}) we then have
\be U(g)\,U^\2O_{L/R}(h)\,U(g)^*=U^\2O_{L/R}(ghg^{-1})=U^\2O_{L/R}(h) ,\ee
that is the disorder operators are gauge invariant and thus contained in 
$\hat{\2A}(\2O)$. It is then obvious that 
\be \hat{\2A}(\2O)=\2A(\2O)\vee U^\2O_L(G)'',\ \ \ \mbox{($G$ abelian!)}\ee
as $\hat{\2A}(\2O)$ is spanned by operators of the form 
$FU^\2O_L(g),\, F\in\2F(\2O)$ which are invariant iff $F\in\2A(\2O)$.

The case of the group $G$ being non-abelian is more complicated and we limit
ourselves to finite groups leading already to structures which are quite 
interesting. In order to proceed we would like to know that every operator
$\hat{F}\in\hat{\2F}(\2O)$ has a  unique representation of the form
\be \hat{F}=\sum_{g\in G} F(g) \, U^\2O_L(g), \ \ \ F(g)\in\2F(\2O) .
\label{repres}\ee
While this true for the crossed product $\2M\rtimes G$ on $L^2(G,\2H)$ (only
for finite groups!) it is not obvious for the algebra $\2M\vee U(G)''$ on $\2H$. 
The latter may be considered as the image of the former under a homomorphism which 
might have a nontrivial kernel. In this case there would be equations of the type
\be \sum_{g\in G} F(g)\, U^\2O_L(g)=0 ,\ee
where not all $F(g)$ vanish. Fortunately at least for finite groups (infinite, thus 
noncompact, discrete groups are ruled out by the split property) this undesirable 
phenomenon can be excluded without imposing further assumptions using the following
result due to Buchholz.
\bprop The automorphisms $\alpha_g=Ad\,U(g)$ act outerly on the wedge algebras.
\label{outer}\eprop
\prf Let $W$ be the standard wedge $W=\{x\in\7R^2 \mid x^1>|x^0|\}$ and assume 
there is a unitary $V_g\in\2F(W)$ such that $Ad\,V_g\restr\2F(W)=\alpha_g$.
Define $V_{g,x}=\alpha_x(V_g)$ for all $x\in W$. Obviously $V_{g,x}\in\2F(W_x)$.
By the commutativity $\alpha_x\circ\alpha_g=\alpha_g\circ\alpha_x$ of translations
and gauge transformations we have $Ad\,V_{g,x}\restr\2F(W_x)=\alpha_g$. By the
computation (for $x\in W$)
\be\ba{ccccccc} V_g\,V_{g,x}\,V_g^* &=& \alpha_g(V_{g,x}) 
   &=& \alpha_g\circ\alpha_x(V_g) &=& \alpha_x\circ\alpha_g(V_g) \\ 
  &=& \alpha_x(V_g\,V_g\,V_g^*) &=& \alpha_x(V_g) &=& V_{g,x} \ea\ee
we obtain
\be V_g\,V_{g,x} = V_{g,x}\,V_g \ \ \forall x\in W .\label{centr}\ee
The von Neumann algebra 
\be \2V=\{V_{g,x},\ x\in W\}'' \ee
is mapped into itself by translations $\alpha_x$ where $x\in W$ and the vacuum 
vector $\Omega$ it is separating for $\2V$ as we have $\2V\subset\2F(W)$. This
allows us to apply the arguments in \cite{dri} to conclude that $\2V$
is either trivial (i.e.\ $\2V=\7C\11$) or a factor of type $III_1$. The assumed
existence of $V_g$, which cannot be proportional to the identity due to the postulate
$\alpha_g\ne\mbox{id}$, excludes the first alternative whereas the second 
is incompatible with (\ref{centr}) according to which $V_g$ is central. Contradiction!
\qed\\
\rem This result may be interpreted as a manifestation of an ultraviolet problem. The 
automorphism $\alpha_g$ being inner on a wedge $W$, wedge duality would imply it to be 
inner on the complementary wedge $W'$, too, giving rise to a factorization 
$U(g)=V_L(g)\,V_R(g),\ V_L(g)\in\2F(W),\,V_R(g)\in\2F(W')$. This would be 
incompatible with the distributional character of the local current from which
$U(g)$ derives.

We cite the following well known result on automorphism groups of factors.
\bprop Let $\2M$ be a factor and $\alpha$ an outer action of the finite group $G$.
Then the inlusions $\2M^G\subset\2M,\ \pi(\2M)\subset\2M\rtimes G$ are irreducible, i.e.\
$\2M\rtimes G\cap\pi(\2M)'=\2M\cap {\2M^G}'=\7C\11$. In particular $\2M\rtimes G$ and 
$\2M^G$ are factors. If the action $\alpha$ is unitarily implemented $\alpha_g=Ad\,U(g)$
then $\2M\rtimes G$ and $\2M\vee U(G)''$ are isomorphic. \label{autom}\eprop
\prf The irreducibility
statements $\2M\rtimes G\cap\pi(\2M)'=\2M\cap {\2M^G}'=\7C\11$ are standard consequences
of the relative commutant theorem \cite[\S 22]{stra} for crossed products.
Remarking that finite groups are discrete and compact the proof is completed by
an application of \cite[Corollary 2.3]{haga} which states that 
$\2M\rtimes G$ and $\2M\vee U(G)''$ are isomorphic if the former algebra 
is factorial and $G$ is compact. \qed

We are now in a position to prove several important corollaries to Proposition
\ref{outer}.
\bcoro The algebras $\2F(\2O),\ \2O\in\2K$ are factors also in the Bose-Fermi case.
\label{factor}\ecoro 
\prf Since $Ad\,V$ acts outerly on the factor $\2F(W^\2O_R)$ by Proposition \ref{outer} 
$M_1=\2F(W^\2O_R)\vee\{V\}$ is a factor and there is an automorphism $\beta$ of 
$M_1$ leaving $\2F(W^\2O_R)$ pointwise invariant such that $\beta(V)=-V$. The 
automorphism $\beta\otimes\alpha_k$ of $M_1\otimes\2F(W^\2O_L)$ clearly has
$Y^\2O\,\2F(\2O)\,Y^{\2O*}$ as fixpoint algebra, cf.\ (\ref{newisom3}). Since
$\alpha_k$ is outer the same holds \cite[Prop. 17.6]{stra} for $\beta\otimes\alpha_k$.
Thus the fixpoint algebra is factorial by another application of Proposition \ref{autom}.
\qed
\bcoro Let $\2O\in\2K$. The automorphisms $\alpha_g=Ad\,U(g)$ and 
$\alpha^\2O_g=Ad\,U^\2O_L(g)$ act outerly on the algebra $\2F(\2O)$. \label{outer2}\ecoro
\prf The pure Bose case is easy. $\2F(\2O),\ \alpha_g^\2O$ and $\alpha_g$ are 
unitarily equivalent to $\2F(W^\2O_R)\otimes\2F(W^\2O_L),\ \ \alpha_g\otimes id$, and
$\alpha_g\otimes\alpha_g$, respectively. Since $\alpha_g=Ad\,U(g)$ is outer on
$\2F(W^\2O_R)$ the same holds by \cite[Prop. 17.6]{stra} for the automorphisms 
$\alpha_g\otimes id$ and $\alpha_g\otimes\alpha_g$ of the above tensor product.

Turning to the Bose-Fermi case let $X_g\in\2F(\2O)$ be an implementer of $\alpha_g$
or $\alpha^\2O_g$ and define $\hat{X}_g=Y^\2O\,X_g\,Y^{\2O*}$. Then 
$(\11\otimes V)\hat{X}_g(\11\otimes V)$  also implements  $\alpha_g\otimes id$ or 
$\alpha_g\otimes\alpha_g$, respectively, since $k$ is central. $\2F(\2O)$ being a factor
this implies 
$(\11\otimes V)\hat{X}_g(\11\otimes V)=c_g\,\hat{X}_g$ with $c_g^2=\pm 1$ due to $k^2=e$.
$\hat{X}_g$ is thus contained either in $\2F(W^\2O_R)\otimes\2F(W^\2O_L)_+$ or in
$\2F(W^\2O_R)\,V\otimes \2F(W^\2O_L)_-$. In the first case the restriction of 
$\alpha_g\otimes id$ or $\alpha_g\otimes\alpha_g$ to $\2F(W^\2O_R)\otimes\2F(W^\2O_L)_+$
is inner which can not be true by the same argument as for the Bose case. (Observe that
$\2F(W^\2O_L)_+$ is factorial.) On the other hand, no
$\hat{X}_g\in\2F(W^\2O_R)\,V\otimes\2F(W^\2O_L)_-$ can implement $\alpha_g\otimes id$ or 
$\alpha_g\otimes\alpha_g$ since both automorphisms are trivial on the subalgebra
$\11\otimes\2F(W^\2O_L)\cap U(G)'$ which requires 
$\hat{X}_g\in\2B(\2H)\otimes{\2F(W^\2O_L)^G}'$. This, however, is impossible: 
$\2F(W^\2O_L)_-\cap{\2F(W^\2O_L)^G}'=[\2F(W^\2O_L)\cap{\2F(W^\2O_L)^G}']_-
=[\7C\11]_-=\emptyset$,
where we have used the irreducibility of $\2F(W^\2O_L)^G\subset\2F(W^\2O_L)$. \qed
\bcoro Let the symmetry group $G$ be finite. Then the enlarged algebra
$\hat{\2F}(\2O)=\2F(\2O)\vee U^\2O_L(G)''$
is isomorphic to the crossed product $\2F(\2O)\rtimes_{\alpha^\2O} G$
and the inclusions $\2A(\2O)\subset\2F(\2O),\ \2F(\2O)\subset\hat{\2F}(\2O)$ are
irreducible.\label{corr1}\ecoro
\prf Obvious from Proposition \ref{autom} and Corollaries 
\ref{factor}, \ref{outer2}. \qed\\
\rem If $G$ is a compact continuous group outerness of the action does not allow
us to draw these conclusions. In this case an additional postulate is needed.
It would be sufficient to assume irreducibility of the inclusion 
$\2A(W)\subset\2F(W)$, for, as shown by Longo, this property in conjunction with
proper infiniteness of $\2A(W)$ implies dominance of the action and factoriality of
the crossed product.

We are now able to give an explicit description of the dual net $\hat{\2A}$.
\btheor Every operator $\hat{A}\in\hat{\2A}(\2O)$ can be uniquely written in the form
\be \hat{A}=\sum_{g\in G} A(g)\, U^\2O_L(g), \label{repres2}\ee
where the $A(g)\in\2F(\2O)$ satisfy
\be A(kgk^{-1})=\alpha_k(A(g)) \ \ \forall g,k\in G.\label{ax2}\ee
Conversely, every choice of $A(g)$ complying with this constraint gives rise to an 
element of $\hat{\2A}(\2O)$. An analogous representation for the algebras
$\hat{\2A}(W^\2O_R)$ is obtained by replacing $U^\2O_L(g)$ by $U(g)$.
\label{ahatstructure}\etheor
\rem Condition (\ref{ax2}) implies $A(g)\in \2F(\2O)\cap U(N_g)'$
where $N_g=\{h\in G\mid gh=hg\}$ is the normalizer of $g$ in $G$. \\
\prf By Proposition \ref{outer} any $\hat{A}\in\hat{\2A}(\2O)$ can be represented
uniquely according to (\ref{repres2}). Since $\alpha_k(\hat{A})$ is given by
$\sum_g \alpha_k(A(g))\, U^\2O_L(kgk^{-1})=\sum_g \alpha_k(A(k^{-1}gk))\, U^\2O_L(g)$
equation (\ref{ax2}) follows by comparing coefficients. It is obvious that the 
arguments can be reversed. The statement on the wedge algebras $\hat{\2A}(W^\2O_R)$ 
follows from the fact that $\hat{\2F}(W^\2O_R)$ is the crossed product of $\2F(W^\2O_R)$
by the global automorphism group, cf. Proposition \ref{prop6}. \qed

\subsection{The Split Property}
The prominent role played by the split property in our investigations so far gives
rise to the question whether it extends to the enlarged nets $\hat{\2A}$ and
$\hat{\2F}$. As to the net $\hat{\2F}$ it is clear that a twist operation is
needed in order to achieve commutativity of the algebras of two spacelike separated 
regions. Let $\2O_1<\2O_2$ be double cones. Then one has 
$\hat{\2F}(\2O_2)^T\subset\hat{\2F}(\2O_1)'$ where
\be \left(\sum_g F(g)\,U^\2O_L(g)\right)^T :=
   \sum_g F(g)^t\,U^\2O_L(g)\ U(g^{-1}) =\sum_g F(g)^t\,U^\2O_R(g)^* ,
\label{twist}\ee
and the $^t$ on $F(g)$ denotes the Bose-Fermi twist of the introduction.
(By the crossed product nature of the algebras $\hat{\2F}(\2O)$ it is clear
that this map is well defined and invertible.) That commutativity holds as claimed
follows easily from $\hat{\2F}(\2O_1)\subset\2F(W^{\2O_1}_L)$ and
$\hat{\2F}(\2O_2)^T\subset\2F(W^{\2O_2}_R)^t$. It is interesting to observe
that the twist has to be applied to the algebra located to the right for this 
construction to work. This twist operation lacks, however, several indispensable
features. Firstly, there is no unitary operator $S$ implementing the twist as in
the Bose-Fermi case. The second, more important objection refers to the fact that
the map (\ref{twist}) becomes non-invertible when extended to right-handed wedge 
regions, for the operators $U^\2O_R(g)$ are contained in $\2F(W^\2O_R)$.

Concerning the net $\hat{\2A}$ which, in contrast, is local there is no conceptual
obstruction to proving the split property. We start by observing that 
$\hat{\2A}(W^\2O_{LL})=\2A(W^\2O_{LL})$. Furthermore, in restriction to a 
simple sector $\2H_1$ wedge duality (Proposition \ref{prop1}) implies
$\hat{\2A}(W^\2O_{RR})\restr\2H_1=\2A(W^\2O_{RR})\restr\2H_1$. As the 
split property for 
the fields carries over \cite{dopl} to the observables in the vacuum sector
there is nothing to do if we restrict ourselves to the latter. We
intend to prove now that the net $\hat{\2A}$ fulfills the split property
on the big Hilbert space $\2H$. To this purpose we draw upon the pioneering work 
\cite{dopl} where it
was shown that the split property (for double cones) of a field net with group 
symmetry and twisted locality follows from the corresponding property of the 
fixpoint net provided the group $G$ is finite abelian. (The case of general groups 
constitutes an open problem, but given nuclearity for the observables and some
restriction on the masses in the charged sectors nuclearity and thus the
split property for the fields can be proved.) 
\bprop The net $\2O\mapsto\hat{\2A}(\2O)$ satisfies the split property for
wedge regions, provided the group $G$ is finite. \label{ahatsplit}\eprop
\prf The split property for wedges is equivalent \cite{bu} to the existence, for
every double cone $\2O$, of a product state $\phi^\2O$ satisfying
$\phi^\2O(AB)=\phi^\2O(A)\cdot\phi^\2O(B)\ 
\forall A\in\hat{\2A}(W^\2O_{LL}), B\in\hat{\2A}(W^\2O_{RR})$. For the rest
of the proof we fix one double cone $\2O$ and omit it in the formulae.
We have already remarked that for the net $\2A$ product states $\phi_0$ are known 
to exist.
In order to construct a product state for $\hat{\2A}$ we suppose $\gamma_e$ is 
a conditional expectation from 
$\2A(W_{LL})\vee\hat{\2A}(W_{RR})$ to $\2A(W_{LL})\vee\2A(W_{RR})$
such that $\gamma_e(\hat{\2A}(W_{RR}))=\2A(W_{RR})$.
Then $\gamma_e(AB)=\gamma_e(A)\,\gamma_e(B)$ where $A,\, B$ are 
as above, implying that $\phi=\phi_0\circ\gamma_e$ is a product state. 
It remains to find the conditional expectation $\gamma_e$.
To make plain the basic idea we consider abelian groups $G$ first. In this
case $\gamma_e$ is given by
\be \gamma_e(\hat{A})= \frac{1}{|G|}\sum_{\chi\in\hat{G}} \psi_\chi^* \hat{A}
  \psi_\chi ,\ee
where $\psi_\chi\in\2F(\2O)$ is a unitary field operator transforming according
to $\alpha_g(\psi_\chi)=\chi(g)\cdot\psi_\chi$ under the group $G$. This map
has all the desired properties. The pointwise invariance of $\hat{\2A}(W_{LL})$ 
follows from the fact that this algebra commutes with the unitaries $\psi_\chi$. On the
other hand 
\be \psi_\chi^*\, U^{\tilde{\2O}}_L(g)\,\psi_\chi=\chi(g)\cdot U^{\tilde{\2O}}_L(g),
   \ \ \ \tilde{\2O}\subset W^\2O_{RR} \ee
in conjunction with the identity $\sum_{\chi\in\hat{G}}\chi(g)=|G|\,\delta_{g,e}$ 
(valid also for non-abelian groups) implies that the 
operators $U^{\tilde{\2O}}_L(g)\in\hat{\2A}(W_{RR}),\ g\ne e$ are annihilated by 
$\gamma_e$. Finally, the existence of $\psi_\chi\in\2F(\2O)$ for all $\chi$ (i.e.\ the 
dominance of
the group action $\alpha$ on $\2F(\2O)$) is well known to follow from the outerness of
the group action $\alpha$. The generalization to non-abelian groups is straightforward.
The unitaries $\psi_\chi$ are replaced by multiplets $\psi_{r,i}$ of isometries
for all irreducible representations $r$ of $G$. They fulfill the following
relations of orthogonality and completeness:
\bea \psi_{r,i}^*\,\psi_{r,j} &=& \delta_{i,j}\11, \label{ortho}\\
 \sum_{i=1}^{d_r}\psi_{r,i}\,\psi_{r,i}^* &=& \11 \label{compl}\eea
and transform according to 
\be \alpha_g(\psi_{r,i})=\sum_{i'} D^r_{i',i}(g)\,\psi_{r,i'} \ee
under the group. That the conditional expectation $\gamma_e$ given by
\be \gamma_e(\hat{A})= \frac{1}{|G|}\sum_{r\in\hat{G}} \sum_{i=1}^{d_r}
 \psi_{r,i}^* \hat{A} \psi_{r,i} ,\ee
does the job follows from
\be \sum_{i=1}^{d_r} \psi_{r,i}^* \,U^{\tilde{\2O}}_L(g)\, \psi_{r,i} = 
   tr\, D^r(g)\cdot U^{\tilde{\2O}}_L(g) =\chi_r(g)\cdot U^{\tilde{\2O}}_L(g).\ee
Again the existence of such multiplets is guaranteed by our assumptions.
\qed\\
\rem Tensor multiplets satisfying (\ref{ortho}, \ref{compl}) were first considered 
in \cite{dr1} where the relation between the charged fields in a net of field 
algebras and the inequivalent representations of the observables was studied
in the framework of \cite{dhr1}. Multiplets of this type will play a role in our
subsequent investigations, too.

\subsection{Irreducibility of $\2A(\2O)\subset\hat{\2F}(\2O)$}
The inclusions $\2A(\2O)\subset\2F(\2O)\subset\hat{\2F}(\2O)$ 
are of the form 
\be \2N=\2P^K\subset\2P\subset \2P\rtimes L=\2M ,\ee
where $K$ and $L$ are finite subgroups of Aut$\,\2P$, as studied in \cite{bisch}
(albeit for type $II_1$ factors). There $\2P^K\subset\2P\rtimes L$ was 
shown to be irreducible iff $K\cap L=\{e\}$ in Out$\,\2P$ and to be of
finite depth if and only if the subgroup $Q$ of Out$\,\2P$ generated by $K$ and $L$ 
is finite. Furthermore, the inclusion has depth two 
(i.e.\ $\2N'\wedge\2M_2$ is a factor where 
$\2N\subset\2M\subset\2M_1\subset\2M_2\subset\cdots$ is the Jones tower 
corresponding to the subfactor $\2N\subset\2M$) in the special case when 
$Q=K\cdot L$ (i.e.\ every $q\in Q$ can be written as $q=kl,\, k\in K, l\in L$). 

In our situation, where $K=\mbox{Diag}(G\times G)$ and $L=G\times \11$,
all these conditions are fulfilled, as we have $Q=G\times G$ and 
$g\times h= (h\times h)\cdot (h^{-1}g\times e)$. The interest of this 
observation for our purposes derives from the following result, discovered by 
Ocneanu and proved, e.g., in \cite{szym2,lo2}. It states that an irreducible 
inclusion $\2N\subset\2M$ arises via 
$\2N=\2M^H=\{x\in\2M\mid\gamma_a(x)=\ve(x)\11\ \forall a\in H\}$ from 
the action of a Hopf algebra $H$ on $\2M$  iff the inclusion has depth two. 
In the next section this Hopf algebra will be identified and related to our quantum
field theoretic setup.

For the irreducibility of $\2A(\2O)$ in $\hat{\2F}(\2O)$ we now give 
a proof independent of any sophisticated inclusion theoretic machinery.
\bprop\label{irred} For any $\2O\in\2K$ we have
\be \hat{\2F}(\2O)\wedge\2A(\2O)'=\7C\11 .\ee\eprop
\prf All unitary equivalences in this proof are implemented by $Y^\2O$. With the 
abbreviations  $\2M_1=\2F(W^\2O_R)^t$ and $\2M_2=\2F(W_L^\2O)$ we have
$\2M_1'\vee\2M_2'\cong\2M_1'\otimes\2M_2'$. By (\ref{isom2}) 
if $\2F$ is bosonic or (\ref{newisom3}) in the Bose-Fermi case we have 
\be \hat{\2F}(\2O)\cong\2F(W^\2O_R)\vee U(G)''\otimes\2F(W^\2O_L)=\2M_1\vee U(G)''
  \otimes\2M_2 ,\ee
where we have used $\2M^t\vee U(G)''=\2M\vee U(G)''$ (which is true for every von 
Neumann algebra $\2M$). Furthermore, 
\bea \2A(\2O)' &=&\2F(\2O)'\vee U(G)'' 
   = (\2F(W^\2O_{LL})\vee\2F(W^\2O_{RR}))^t\vee U(G)'' \\
   &=& \2F(W^\2O_{LL})\vee\2F(W^\2O_{RR})\vee U(G)'' 
   = \2F(W^\2O_{LL})\vee\2F(W^\2O_{RR})^t\vee U(G)'' \nn\\
   &=& \2M_1'\vee\2M_2'\vee U(G)'' 
 \cong (\2M_1'\otimes\2M_2')\vee\{U(g)\otimes U(g),\, g\in G \}'' .\nn\eea
The relative commutant $\hat{\2F}(\2O)\wedge\2A(\2O)'$ is thus equivalent to
\be (\2M_1\vee U(G)'' \otimes\2M_2) \wedge  [(\2M_1'\otimes\2M_2') 
   \vee \{U(g)\otimes U(g),\, g\in G \}''] .\label{relcomm}\ee
The obvious inclusion 
$(\2M_1'\otimes\2M_2') \vee \{U(g)\otimes U(g),\, g\in G \}''\subset
\2B(\2H)\otimes\2M_2'\vee U(G)''$ 
in conjunction with the irreducibility property
$\2M_2\wedge(\2M_2'\vee U(G)'')=\7C\11$ (Corollary \ref{corr1}) yields
\be [(\2M_1'\otimes\2M_2') \vee \{U(g)\otimes U(g),\, g\in G \}'']\wedge
(\2B(\2H)\otimes\2M_2)\subset\2B(\2H)\otimes\11 .\ee
Now let $X$ be an element of the algebra given by eq. (\ref{relcomm}).
By the same arguments as used earlier, every operator
$X\in(\2M_1'\otimes\2M_2') \vee \{U(g)\otimes U(g),\, g\in G \}''$ has a
unique representation of the form $X=\sum_g F_g\, (U(g)\otimes U(g))$ where
$F_g\in\2M_1'\otimes\2M_2'$. The condition $X\in\2B(\2H)\otimes\11$ implies
$F_g=0$ for all $g\ne e$ and thereby $X\in\2M_1'\otimes\11$. We thus have
$X\in(\2M_1'\wedge(\2M_1\vee U(G)''))\otimes\11$ and, once again using the
irreducibility of the group inclusions, $X\propto\11\otimes\11$. \qed

\sectreset{Quantum Double Symmetry}
\subsection{Abelian Groups}
As we have shown above the algebras $\hat{\2F}(\2O)$ may be considered as
crossed products of $\2F(\2O)$ with the actions of the respective automorphism
groups $\alpha^\2O$. In the case of abelian (locally compact) groups there is a
canonical action \cite{takes} of the dual (character-) group $\hat{G}$ on
$\2M\rtimes G$ given by
\be\ba{ccc} \hat{\alpha}_\chi (\pi(x)) & = & \pi(x) \\
  \hat{\alpha}_\chi(U_g) & = & \ol{\chi(g)} \cdot U_g\ea\ \ ,\chi\in\hat{G}.
\label{alphahat}\ee
Making use of $U^{\2O_1}(g)\,U^{\2O_2}(g)^*\in \2F,\ \forall\2O_i$ 
one can consistently define an action of $\hat{G}$ on the net 
$\2O\mapsto\hat{\2F}(\2O)$, respecting the local structure and thus extending
to the quasilocal algebra $\hat{\2F}$. The action of 
$\hat{G}$ commutes with the original action of $G$ as extended to $\hat{\2F}$, 
implying that the locally compact group $G\times\hat{G}$ is a group of local
symmetries of the extended theory $\2O\mapsto\hat{\2F}(\2O)$. The square
structure (\ref{square}) can now easily be interpreted in terms of the larger
symmetry:
\be \hat{\2A}=\hat{\2F}^G,\ \ \2F=\hat{\2F}^{\hat{G}},\ \ 
   \2A=\hat{\2F}^{G\times\hat{G}} .\ee
The symmetry between the subgroups $G$ and $\hat{G}$ of $G\times\hat{G}$ is, however,
not perfect, as only the automorphisms $\alpha_g,\, g\in G$ are unitarily implemented
on the Hilbert space $\2H$. That there can be no unitary implementer $U(\chi)$ for
$\hat{\alpha}_\chi,\,\chi\in\hat{G}$ leaving invariant the vacuum $\Omega$ is shown 
by the following computation which would be valid for all $A\in\2A(\2O)$
\bea \lefteqn{\langle\Omega,AU^\2O_L(g)\Omega\rangle
   =\langle\Omega,U(\chi)\,AU^\2O_L(g)\,U(\chi)^*\Omega\rangle} \\
  && =\langle\Omega,A\hat{\alpha}_\chi(U^\2O_L(g))\Omega\rangle
  =\ol{\chi(g)}\cdot \langle\Omega,AU^\2O_L(g)\Omega\rangle .\nn\eea
This can only be true if $\chi(g)=1$ or 
$\langle\Omega,AU^\2O_L(g)\Omega\rangle=0 \ \forall A\in\2A(\2O)$.
The latter, however, can be ruled out, since the density of $\2A(\2O)\Omega$ in $\2H_0$ 
would imply $U^\2O_L(g)\Omega\perp\2H_0$ which is impossible, $\Omega$ being unitary and
gauge invariant. This argument shows that the vacuum state 
$\omega=\langle\Omega, \cdot\,\Omega\rangle$ is not invariant under the automorphisms
$\hat{\alpha}(\chi),\ \chi\in\hat{G}$, in other words, the symmetry under $\hat{G}$
is spontaneously broken. 

The preceding argument is just a special case of the much
more general analysis in \cite{rob}, where non-abelian groups were considered, too.
There, to be sure, the field net acted upon by the group was supposed to fulfill
Bose-Fermi commutation relations, whereas in our case the field net is nonlocal.
Furthermore, whereas the net $\2F(\2O)$, the point of departure for our analysis,
fulfills (twisted) duality, the extended net $\hat{\2F}(\2O)$ enjoys no obvious 
duality properties. Nevertheless the analogy to \cite{rob} goes beyond the above
argument. Indeed, as shown by Roberts, spontaneous breakdown of group symmetries
is accompanied by a violation of Haag duality for the observables, restricted to
the vacuum sector $\2H_0$. Defining the net $\2B(\2O)=\2F(\2O)^{G_0}$, 
the fixpoint net under the action of the unbroken part 
$G_0=\{g\in G\,|\,\omega_0\circ\alpha_g=\omega_0\}$ of the symmetry group,
a combination of the arguments in \cite{dhr1} and \cite{rob} leads to the conclusion
that (in the vacuum sector $\2H_0$) $\2B(\2O)$ is just the dual net 
$\2A^d(\2O)$ which verifies Haag duality. Our analysis in Section 2, leading to 
the identification of the dual net as $\2A^d=\hat{\2A}=\hat{\2F}^G$, is 
obviously in accord with the general theory as we have shown above that $G$ 
is the unbroken part, corresponding to $G_0$, of the full symmetry group 
$G\times\hat{G}$.

In the case of spontaneously broken group symmetries it is known that, irrespective
of the nonexistence of global unitary implementers leaving invariant the vacuum,
one can find local implementers for the whole symmetry group. This means that for
each double cone $\2O$ there exists a unitary representation 
$G\ni g\mapsto V_\2O(g)$ satisfying $Ad\,V_\2O(g)\restr\2F(\2O)=\alpha_g$,
the important point being the dependence on the region $\2O$. (Due to the large
commutant of $\2F(\2O)$ such operators are far from unique.)
A particularly nice construction, which applied to an unbroken symmetry $g$ 
automatically yields the global implementer ($V_\2O(g)=U(g)\,\forall\2O$), 
was given in \cite{bdlr}. The construction
given there applies without change to the situation at hand where the action of the 
dual group $\hat{G}$ on $\hat{\2F}(\2O)$ is spontaneously broken. 

An interesting example is provided by the free massive Dirac field which
as already mentioned fulfills our postulates, including twisted duality and the 
split property. Its symmetry group $U(1)$ being compact and abelian, the extended
net $\hat{\2F}$ and the action of the dual group $\7Z$ can be constructed as
described above. By restriction of the net $\hat{\2A}$ to the vacuum sector 
$\2H_0$ one obtains a local net fulfilling Haag duality with symmetry group 
$\7Z$. Wondering to which quantum field theory this net might correspond, it 
appears quite natural to think of the sine-Gordon theory at the free fermion point
$\beta^2=4\pi$ as discussed, e.g., in \cite{lehm}. 

\subsection{Non-Abelian Groups}\label{o-nonabel}
We refrain from a further discussion of the abelian case and turn over to the
more interesting case of $G$ being non-abelian and finite. (Infinite compact groups
will be treated in Appendix B.) For non-abelian
groups the dual object is not a group but either some Hopf algebraic structure or a 
category of representations. Correspondingly, the action of the dual group in 
\cite{takes} has to be replaced by a coaction of the group or the action of a
group dual in the sense of \cite{rob2}. For our present purposes these high-brow 
approaches
will not be necessary. Instead we choose to generalize (\ref{alphahat}) in the
following straightforward way. We observe that the characters of a compact abelian 
group constitute
an orthogonal basis of the function space $L^2(G)$, whereas in the non-abelian case
they span only the subspace of class functions. This motivates us to define
an action of $\7C(G)$, the $|G|$-dimensional space of {\it all} complex valued 
functions on $G$, on $\hat{\2F}(\2O)$ in the following way:
\be \gamma_F\left(\sum_{g\in G} x(g)\,U^\2O_L(g)\right)= \sum_{g\in G} 
   F(g)\,x(g)\,U^\2O_L(g) ,\ \ x(g)\in\2F(\2O), F\in \7C(G). \label{gamma_F}\ee
Again this action of $\7C(G)$ is consistent with the local structure of the net 
$\2O\mapsto\hat{\2F}(\2O)$ and extends to the quasilocal $C^*$-algebra $\hat{\2F}$.
In general, of course, $\gamma_F$ is no homomorphism but only a linear map. 
(That the maps $\gamma_F$ are well defined for every $F\in\7C(G)$ should be 
obvious, see also the next section.)
Introducing the `deltafunctions' $\delta_g(h)=\delta_{g,h}$ any function can be 
written as $F=\sum_g F(g)\,\delta_g$, and $\gamma_{\delta_g}$ will be abbreviated
by $\gamma_g$. The latter are projections, i.e.\ they satisfy $\gamma_g^2=\gamma_g$.
The images of $\hat{\2F}(\2O)$ and $\hat{\2F}$ under these will be 
designated $\hat{\2F}_g(\2O)$ and $\hat{\2F}_g$, respectively. Obviously
we have $\hat{\2F}_g(\2O)=\2F(\2O)\,U_L^\2O(g)$ and
$\hat{\2F}_g=\2F\,U_L^\2O(g)$ with $\2O\in\2K$ arbitrary.
It should be clear that the decomposition
\be \hat{\2F}=\bigoplus_{g\in G}\hat{\2F}_g \ee
represents a grading of $\hat{\2F}$ by the group, i.e.
\be \hat{\2F}_g\,\hat{\2F}_h\subset\hat{\2F}_{gh} \ \forall g,h\in G .\ee
(In fact we have equality, but this will play no role in the sequel.)
This group grading which is, of course, not surprising as it holds for every crossed
product by a finite group allows us to state the behavior of $\gamma_g$ under products:
\be \gamma_g(AB)=\sum_h \gamma_h(A)\,\gamma_{h^{-1}g}(B) .\label{grad2}\ee
The novel aspect, however, is that $\hat{\2F}$ is
at the same time acted upon by the group $G$, these two structures being coupled by
\be \alpha_g(\hat{\2F}_h)=\hat{\2F}_{ghg^{-1}} \label{coupl}\ee
as a consequence of (\ref{cov}). This is equivalent to the relation
\be \alpha_g\circ\gamma_h=\gamma_{ghg^{-1}}\circ\alpha_g .\label{cov2}\ee
In this context it is of interest to remark that several years ago algebraists
studied (see \cite{cohen} and references given there) analogies between group 
graded algebras and algebras acted upon by a finite group. Similar studies have been
undertaken in the context of inclusions of von Neumann algebras. As it turns out 
the situation at hand, which is rather more interesting, can be neatly described 
in terms of the action, as defined, e.g., in \cite{szym1}, of a Hopf algebra 
(in our case finite 
dimensional) on $\hat{\2F}$. The relations fulfilled by the $\alpha_g$ and 
$\gamma_h$, in particular (\ref{cov2}), motivate us to cite the following
well known
\bdefin Let $\7C(G)$ be the algebra  of (complex valued) functions on the finite
group $G$ and consider the adjoint action of $G$ on $\7C(G)$ according to 
$\alpha_g:\ f\mapsto f\circ Ad(g^{-1})$. The quantum double $D(G)$ is defined 
as the crossed product $D(G)=\7C(G)\rtimes_\alpha G$ of $\7C(G)$ by this
action. In terms of generators $D(G)$ is the algebra 
generated by elements $U_g$ and $V_h,\, g,h\in G$ with the relations
\bea U_g\,U_h &=& U_{gh} \\
     V_g\,V_h &=& \delta_{g,h}V_g \\
     U_g\,V_h &=& V_{ghg^{-1}}\,U_g  \eea
and the identification $U_e=\sum_g V_g=\11$. 
\label{double1}\edefin
It is easy to see that $D(G)$ is of the finite dimension $|G|^2$, where as a
convenient basis one may choose $V(g)U(h),\,g,h\in G$, multiplying according to
$V(g_1)U(h_1)\, V(g_2)U(h_2)=\delta_{g_1,h_1 g_2 h_1^{-1}}\cdot V(g_1)U(h_1 h_2)$.
This is just a special case 
of a construction given by Drinfel'd \cite{drin1} in greater generality which we do 
not bother to retain. For the purposes of this work it suffices to state the following
well known properties of $D(G)$, referring to \cite{drin1,reshet,dpr} for further 
discussion, see also Appendix A.

In order to define an action of a Hopf algebra on von Neumann algebras we further 
need a star structure on the former which in our case is provided by the following
\bprop With the definition $U_g^*=U_{g^{-1}},\ V_h^*=V_h$ and the appropriate
extension, $D(G)$ is a *-algebra. D(G) is semisimple. \eprop
\prf Trivial calculation. Finite dimensional *-algebras are automatically semisimple.
\qed

Before stating how the quantum double $D(G)$ acts on $\hat{\2F}$ we define
precisely the properties of a Hopf algebra action.
\bdefin\label{def:action}
A bilinear map $\gamma: H\times\2M\to\2M$ is an action of the Hopf *-algebra $H$
on the *-algebra $\2M$ iff the following hold for any $a,b\in H,\ x,y\in\2M$:
\bea
\gamma_\11(x)  &=& x ,\label{action1}\\
\gamma_a (\11) &=& \ve(a)\11 ,\label{action2}\\
\gamma_{ab}(x) &=& \gamma_a \circ \gamma_b(x) ,\label{action3}\\
\gamma_a(xy) &=& \gamma_{a^{(1)}}(x)\gamma_{a^{(2)}}(y) ,\label{action4}\\
(\gamma_a(x))^*  &=& \gamma_{S(a^*)}(x^*) .\label{action5}\eea
We have used the standard notation $\Delta(a)=a^{(1)}\otimes a^{(2)}$ for
the coproduct where on the right side there is an implicit summation.
The map $\gamma$ is assumed to be weakly continuous with respect to
$\2M$ and continuous with respect to some $C^*$-norm on $H$ (which is unique in
the case of finite dimensionality).
\edefin
%\rem Alternatively, one can define an action of $H$ on $\2M$ to be given by a 
%*-homo\-morphism $\delta:\2M\to H\otimes\2M$ satisfying 
%$(\delta_H\otimes\mbox{id})\circ\delta=(\mbox{id}\otimes\delta)\circ\delta$, where
%$\delta_H$ is the coproduct of $H$. This notion is more convenient when studying
%actions of (infinite dimensional) Hopf von Neumann algebras.

After these lengthy preparations it is clear how to define the action of $D(G)$ on
$\hat{\2F}$. 
\btheor Defining 
$\gamma_a(\hat{F}),\ \hat{F}\in\hat{\2F}$ for $a\in\{U(g),\,V(h)|g,h\in G\}$ by
\bea \gamma_{U_g}(\hat{F}) &=& \alpha_g(\hat{F}) \\
  \gamma_{V_h}(\hat{F}) &=& \gamma_h(\hat{F}) ,\eea
using (\ref{action3}) to define $\gamma$ on the basis $V(g)U(h)$ and extending 
linearly to $D(G)$ one obtains an action in the sense of Definition \ref{def:action}.
\label{thm:action}\etheor
\prf (\ref{action1}) follows from $\11_{D(G)}=\sum_g V_g$, (\ref{action2}) from 
$\11_{\hat{\2F}}\in\hat{\2F}_e$ and (\ref{counit}), whereas (\ref{action3}) 
is an obvious consequence of the definition. Furthermore, (\ref{action4}) is a
consequence of $\alpha_g$ being a homomorphism, the coproduct property (\ref{grad2}) 
and the definition (\ref{coprod}). The statement (\ref{action5}) on the 
*-operation finally follows from
$(\alpha_g(x))^*=\alpha_g(x^*)$ and $S(U_g^*)=U_g$ on the one hand and 
$(\hat{\2F}_g)^*=\hat{\2F}_{g^{-1}}$ and $S(V_g^*)=V_{g^{-1}}$ on the other.
\qed\\
\rems 1. It should be obvious that the action of $D(G)$ on $\hat{\2F}$ commutes
with the translations and that it commutes with the boosts iff the group $G$ does.
Otherwise, $U(\Lambda)\,U(g)\,U(\Lambda)^*=U_h$ implies 
$\alpha_\Lambda\circ\gamma_g=\gamma_h\circ\alpha_\Lambda$.\\
2. In the case of $G$ being abelian 
$U_\chi=\sum_{g\in G} \ol{\chi(g)}\cdot V_g,\ \chi\in\hat{G}$ constitutes
an alternative basis for the subalgebra $\7C(G)\subset D(G)$. The resulting
formulae $U_\chi\,U_\rho=U_{\chi\rho},\ \Delta(U_\chi)=U_\chi\otimes U_\chi$ and 
$\gamma_{U_\chi}(\cdot)=\hat{\alpha}_\chi(\cdot)$ establish the equivalence of
the quantum double with the group $G\times\hat{G}$. The abelian case is special
insofar as $D(G)$ is spanned by its grouplike elements, which is not true for
$G$ non-abelian.

\subsection{Spontaneously Broken Quantum Symmetry}
Having shown in the abelian case that the symmetry under the dual group $\hat{G}$
is spontaneously broken it should not come as a surprise that the same holds for
non-abelian groups $G$ where, of course, the notion of unitary implementation has
to be generalized.
\bdefin An action $\gamma$ of the Hopf algebra $H$ on the *-algebra $\2M$ 
is said to be implemented by the (homomorphic) representation
$U:\,H\to\2B(\2H)$ if for all $a\in H, x\in\2M$
\be U(a) \, x = \gamma_{a^{(1)}}(x) \, U(a^{(2)}) \ee
or equivalently
\be \gamma_a(x)= U(a^{(1)}) \, x U(S(a^{(2)})) .\ee
The representation is said to be unitary if the map $U$ is a *-homomorphism.
\edefin
In complete analogy to the abelian case we see that only a subalgebra of $D(G)$, 
namely the group algebra $\7CG$ is implemented in the above sense. A similar
phenomenon has already been observed to occur in the Coulomb gas representation
of the minimal models \cite{gs} and in \cite{bern} where two dimensional theories
without conformal covariance were considered. It would be interesting to know
whether there exists, in some sense, a `quantum version' of Goldstone's theorem for
spontaneously broken Hopf algebra symmetries.

In an earlier section we defined a twist operation (\ref{twist}) which bijectively
maps $\hat{\2F}(\2O)$ into an algebra $\hat{\2F}(\2O)^T$ which commutes with all
field operators localized in the left spacelike complement $W^\2O_{LL}$ of $\2O$.
With the notation introduced in this chapter this operation can be written as
$F^T=\sum_g \gamma_g(F)^t\,U(g^{-1})$. One might wonder whether there is a map 
$\bar{T}$ which achieves the same thing for the right spacelike complement 
$W^\2O_{RR}$. If the quantum symmetry were not spontaneously broken, such a map would 
be given by
\be F^{\bar{T}}=\sum_g \alpha_g(F)^t\, V(g) ,\ee
where the $V(g)$ are the projectors implementing the dual $\7C(G)$ of the group $G$.
Using the spacelike commutation relations and the property 
$U^\2O(g)\,V(h)=V(gh)\,U^\2O(g)$ this claim is easily verified. 

In the discussion of the abelian case we have mentioned that one can construct, e.g.\
by the method given in \cite{bdlr}, local implementers of the dual group $\hat{G}$.
For the quantum double $D(G)$ of a non-abelian group $G$, however,
which is not spanned by its grouplike elements, another approach is needed.
\bprop For every double cone $\2O\in\2K$ there is a family of orthogonal 
projections $V_\2O(g)$ fulfilling
\be V_\2O(g)\,V_\2O(h)=\delta_{g,h}\,V_\2O(g)\ ,\ \ \sum_g\,V_\2O(g)=\11 ,
  \label{v1}\ee
\be \gamma_g\restr\hat{\2F}(\2O)=\sum_h V_\2O(gh)\,\cdot\,V_\2O(h)
  \label{v2}\ee
and transforming correctly under the (unbroken) group $G$
\be U(g)\,V_\2O(h)\,U(g)^* =V_\2O(ghg^{-1}) .\label{v3}\ee
\eprop
\prf 
In order to obtain operators with these properties we make use of the isomorphism,
for every wedge $W$, between $\2F(W)\vee U(G)''$ and $\2F(W)\rtimes_\alpha G$. 
We shortly remind the construction of the crossed product $\2M\rtimes G$. 
It is represented on the Hilbert space $\bar{\2H}=L^2(G,\2H)$ of square 
integrable functions from $G$ to $\2H$. The algebra $\2M$ acts according to 
$(\pi(x)f)(g)=\alpha_{g^{-1}}(x)\,f(g)$
whereas the group $G$ is unitarily represented by $(\bar{U}(k)f)(g)=f(k^{-1}g)$.
With these definitions one can easily verify the equation
$\bar{U}(k)\,\pi(x)\,\bar{U}(k)^*=\pi\circ\alpha_k(x)$.
If the group $G$ is finite one can furthermore define the projections
$(\bar{E}(k)f)(g)=\delta_{g,k}\,f(g)$ for which one obviously has
$\bar{U}(g)\,\bar{E}(k)=\bar{E}(gk)\,\bar{U}(g)$. As already discussed above there is,
as a consequence of the outerness of the action of the group, an isomorphism between 
the algebras 
$\2M\vee U(G)''$ and $\2M\rtimes_\alpha G$ sending $\sum_g x_g\,U(g)$ to
$\sum_g \pi(x_g)\,\bar{U}(g)$. As both algebras are of type III and live on
separable Hilbert spaces this isomorphism is unitarily implemented and can be
used to pull back the projections $\bar{E}(k)$ to the Hilbert space $\2H$
where we denote them by $E(k)$. ($E(e)$ is nothing but the Jones projection in 
the extension $\2M_2$ of the inclusion $\2M\subset\2M\vee U(G)''$.)
Applying these considerations to the algebras of the wedges $W^\2O_L$ and 
$W^\2O_R$ we obtain the families of projections $E^\2O_{L/R}(k)$,
satisfying 
\be U(g)\,E^\2O_{L/R}(k)\,U(g)^*=E^\2O_{L/R}(gk) ,\label{jones}\ee
which we use to
define
\be V_\2O(g)=Y^{\2O *}\,( \sum_h E^\2O_R(gh)\otimes E^\2O_L(h))
   \,Y^\2O .\ee
The properties (\ref{v1}) of orthogonality and completeness are obvious whereas
covariance (\ref{v3}) follows from (\ref{jones}) and 
$U(k)=Y^{\2O*}\,U(k)\otimes U(k)\,Y^\2O$ as follows
\bea Ad \, U(k)(V_\2O(g)) &=& Y^{\2O *}\,( \sum_h 
  E^\2O_R(kgh)\otimes E^\2O_L(kh) )  \,Y^\2O \nn\\
  &=& Y^{\2O *}\,( \sum_h E^\2O_R(kgk^{-1}h)\otimes E^\2O_L(h)) \,Y^\2O\\
  &=& V_\2O(kgk^{-1})  .\nn\eea
It remains to show the implementation property (\ref{v2}). Using
$E^\2O_L(g)\,\2F(W^\2O_L)\,E^\2O_L(h)=\{0\}$ if $g\ne h$ and
$\hat{\2F}(\2O)\cong\2F(W^\2O_R)\vee U(G)''\otimes\2F(W^\2O_L)$ we obtain
\bea Y^\2O\,\sum_h V_\2O(gh)\,\hat{F}\,V_\2O(h)\,Y^{\2O*}
  &=& \sum_{h,k,l} E^\2O_R(ghk)\otimes E^\2O_L(k)\, F_1\otimes F_2\,
     E^\2O_R(hl)\otimes E^\2O_L(l) \nn\\
  &=& \sum_{h,k} E^\2O_R(ghk)\otimes E^\2O_L(k)\, F_1\otimes F_2\,
     E^\2O_R(hk)\otimes E^\2O_L(k) \nn\\
  &=& (\sum_h E^\2O_R(gh)\, F_1\,E^\2O_R(h)) \otimes (\sum_k
     E^\2O_L(k)\, F_2\,E^\2O_L(k)) \nn\\
  &=& \sum_h E^\2O_R(gh)\, F_1\,E^\2O_R(h) \otimes F_2 \eea
where we have written (abusively) $F_1\otimes F_2$ for $Y^\2O\,\hat{F}\,Y^{\2O*}$.
With $\sum_h E^\2O_R(gh)\, U(k)\,E^\2O_R(h)=\delta_{g,k} U(k)$ it is clear that
the above map projects $\hat{\2F}(\2O)$ onto $\2F(\2O)U^\2O_L(g)$,
thus implementing the restriction of $\gamma_g$ to $\hat{\2F}(\2O)$. \qed\\
\rem It should be remarked that the  simpler definition 
$\tilde{V}_\2O(g)=Y^{\2O*}(E^\2O_R(g)\otimes \11)Y^\2O$, which also 
satisfies (\ref{v2}), does not lead to a representation of $D(G)$ as these $V_\2O$'s
do not transform according to the adjoint representation (\ref{v3}).

\subsection{Spectral Properties}
The above discussion was to a large extent independent of the quantum field theoretic 
application insofar as the action of the quantum double on a certain class of 
*-algebras was concerned. As we have seen, any *-algebra which is at the same time 
acted upon by a finite group $G$ and graded by $G$ supports an action of the
double provided the relation (\ref{coupl}) holds. The converse is also true. Let
$\2M$ be a *-algebra on which the double acts. Then $\2M_g=\gamma_g(\2M)$ 
induces a $G$-grading satisfying (\ref{coupl}). It may however happen that 
$\2M_g=\{0\}$ for $g$ in a normal subgroup. This possibility can be eliminated by
demanding the existence of a unitary representation of $G$ in 
$\2M:\ G\ni g\mapsto U(g)\in\2M_g$. In the situation at hand this condition
is fulfilled by construction.

We now turn to the spectral properties of the action of the double. To this 
purpose we introduce the following notion \cite{rob2}, already encountered 
implicitly in the proof of Proposition \ref{ahatsplit}.
\bdefin A normclosed linear subspace $\2T$ of a von Neumann algebra $\2M$
is called a Hilbert space in $\2M$ if $x^*x\in \7C\11$ for all $x\in \2T$ 
and $x\in\2M$ and $xa=0\ \forall a\in\2T$ implies $x=0$. \edefin
The name is justified as $\langle x,y\rangle \11=x^* y$ defines a 
scalar product in $\2T$. One can thus choose a basis $\psi_i,\ i=1,\ldots, d_\2T$ 
satisfying the requirements (\ref{ortho}, \ref{compl}). The interest of this 
definition stems from the following well known lemma, the easy proof of which we omit.
\blemma Let $\2T$ be a finite dimensional Hilbert space in $\2M$ globally 
invariant under the action $\gamma_H$ of $H$ on $\2M$. A basis of the above 
type gives rise to a unitary representation of $H$ according to
\label{tensor}\elemma
\be \gamma_a(\psi_i)=\sum_{i'=1}^d D_{i'i}(a) \, \psi_{i'} .\label{o-gamma_a}\ee

Our aim will now be to show that the extended algebras 
$\hat{\2F}(\2O),\ \2O\in\2K$ in fact contain such tensor multiplets for every 
irreducible representation of $D(G)$. In order to do this we  
make use of the representation theory of the double developed in \cite{dpr}.
($D(G)$ being semisimple, every finite dimensional representation decomposes into 
a direct sum of irreducible ones.) The (equivalence classes of)
irreducible representations are labeled by
pairs $(c, \pi)$, where $c\in C(G)$ is a conjugacy class and $\pi$ is an irreducible
representation of the normalizer group $N_c$. Here $N_c$ is the abstract group 
corresponding to the mutually isomorphic normalizers $N_g$ for $g\in c$, already 
encountered in Theorem \ref{ahatstructure}. 
The representation $\hat{\pi}$ labeled by $(c,\pi)$ is obtained by choosing an 
arbitrary $g_0\in c$ and inducing up from the representation 
\be \hat{\pi}(V_g\,U_h)=\delta_{g, g_0}\,\pi(h) \ee
of the subalgebra $\2B_{g_0}$ of $D(G)$ 
generated by $V(g)\,,g\in G$ and $U(h)\,,h\in N_{g_0}$.
The representation space of $\hat{\pi}_{(c,\pi)}$ is thus
$V_{(c,\pi)}=D(G)\otimes_{\2B_{g_0}} V_\pi$. For a more complete discussion
we refer to \cite{dpr} remarking only that 
$\hat{\pi}_{(c,\pi)}(V_g\,U_h)=0$ if $g\not\in c$.

\bdefin The action $\gamma$ of a group or Hopf algebra on a von Neumann algebra
$\2M$ is dominant iff the algebra of fixed points is properly infinite and the monoidal
spectrum of $\gamma$ is complete, i.e.\ for every finite dimensional unitary 
representation $\pi$ of the group or Hopf algebra, respectively, there is a 
$\gamma$-invariant Hilbert space $\2T$ in $\2M$ such that $\gamma\restr\2T$ is equivalent
to $\pi$. \edefin
\bprop Let $\hat{\2M}$ be a von Neumann algebra supporting an action of the 
quantum double $D(G)$. Assume further that there is a unitary representation of $G$ in
$\hat{\2M}$ where $\bar{U}(g)\in\hat{\2M}_g$ and 
$\alpha_h(\bar{U}(g))=\bar{U}(hgh^{-1})$.
Then the action of $D(G)$ on $\hat{\2M}$ is dominant if and only if the action of $G$ 
on $\2M=\gamma_e(\hat{\2M})$ is dominant. \eprop
\prf As a consequence of $\2M^G=\hat{\2M}^{D(G)}$ the conditions of proper 
infiniteness of the fixpoint algebras coincide. The `only if' statement is easily
seen by considering the representations of the double corresponding to the
conjugacy class $c=\{e\}$. For these $N_c\cong G$ holds, implying that the 
representations of $D(G)$ with $c=\{e\}$ are in one-to-one correspondence to
the representations of $G$. A multiplet in $\hat{\2M}$ transforming according 
to $(\{e\},\pi)$ is nothing but a $\pi$-multiplet in $\2M$.

The `if' statement requires more work. We have to show that for every pair $(c,\pi)$,
where $\pi$ is an irreducible representation of the normalizer $N_c$, there exists
a multiplet of isometries transforming according to $\hat{\pi}_{(c,\pi)}$. To begin 
with, choose $g\in c$ arbitrarily and find in $\2M$ a multiplet of isometries 
$\psi_i,\ i=1,\ldots, d=\dim(\pi)$ transforming according to the representation $\pi$ 
under the action of $N_g\subset G$. The existence of such a multiplet follows from
the dominance of the group action on $\2M$. Now, let $x_1,\ldots, x_n$ be representatives
of the cosets $G/N_g$ where $n=[G:N_g]=|c|$. Furthermore, the proper infiniteness of
the fixpoint algebra allows us to choose a family 
$V_1,\ldots, V_n$ of isometries in $\2M^G=\hat{\2M}^{D(G)}$ satisfying
$V_i^*V_j=\delta_{i,j},\ \sum_i V_i V_i^*=\11$. Defining
\be \Psi_{ij}=V_i\,\alpha_{x_i}(\bar{U}(g)\,\psi_j),\ \ i=1,\ldots, n,\,j=1,\ldots, d \ee
one verifies that the $\Psi_{ij}$ constitute a complete family of mutually
orthogonal isometries spanning a vectorspace of dimension 
$nd=\dim(\hat{\pi}_{(c,\pi)})$.
That this space is mapped into itself by the action of the double follows from the fact
that, for every $k\in G$, $k\,x_i$ can uniquely be written as $x_j\,h,\ h\in N_g$. 
Finally, the multiplet transforms according
to the representation $(c,\pi)$ of $D(G)$, which is evident from the definition of the
latter in \cite[(2.2.2)]{dpr}.\qed\\
\rem Since in our field theoretic application the conditions of the proposition are 
satisfied thanks to Lemma \ref{outer2} and the discussion in Subsection \ref{o-nonabel} 
we can conclude that $\hat{\2F}(\2O),\ \2O\in\2K$ has full $D(G)$-spectrum.

\subsection{Commutation Relations and Statistics}
Up to this point our investigations in this section have focused on the local
inclusion $\2A(\2O)\subset\hat{\2F}(\2O)$ for any fixed double cone $\2O$.
Having clarified the relation between these algebras in terms of the action of 
the quantum double we can now complete our discussion of the latter. 
To this purpose we recall that the double construction has been introduced in 
\cite{drin1} as a means of obtaining quasitriangular Hopf algebras (quantum groups)
in the sense defined there, i.e.\ Hopf algebras possessing a `universal R-matrix',
cf.\ Appendix A.
As it turns out the latter appears very naturally in our approach when considering 
the spacelike commutation relations of irreducible $D(G)$-multiplets as defined in the
preceding subsection.
\bprop Assume the net $\2O\mapsto\2F(\2O)$ is bosonic, i.e.\ fulfills
untwisted locality. Let $\2O_2<\2O_1$ and $\psi^1_i, i=1,\ldots, d_1$ and
$\psi^2_j, j=1,\ldots, d_2$ be $D(G)$-tensors in $\hat{\2F}(\2O_1),\,\hat{\2F}(\2O_2)$, 
respectively. They then fulfill C-number commutation relations
\be \psi_i^1\,\psi_j^2 =\sum_{i'j'} \psi^2_{j'}\,\psi^1_{i'}\ 
   (D^1_{i'i}\otimes D^2_{j'j}) (R) ,\label{commrel3}\ee
where $D^1,\,D^2$ are the matrices of the respective representations of $D(G)$ and 
\be R=\sum_{g\in G} V_g\otimes U_g\in D(G)\otimes D(G) .\label{rmatrix}\ee \eprop
\prf The equation $\sum_g V_g=\11$ in $D(G)$ implies $\sum_g \gamma_g=id$. We can thus
compute 
\bea \psi_i^1\,\psi_j^2 &=& \sum_{g\in G} \gamma_g(\psi_i^1)\,\psi_j^2 =
  \sum_{g\in G} \alpha_g(\psi_j^2)\, \gamma_g(\psi_i^1)\\
  &=& \sum_{g\in G} \sum_{i'j'} \psi^2_{j'}\,\psi^1_{i'}\
     D^2_{j'j}(U_g)\,D^1_{i'i}(V_g) ,\nn\eea
where the second identity follows from $\gamma_g(\psi_i^1)\in\2F(\2O_1)\,U^{\2O_1}_L(g)$
and $Ad\,U^{\2O_1}_L(g)\restr\hat{\2F}(\2O_2)=\alpha_g$. The 
rest is clear. \qed\\
\rems 1. Commutation relations of the above general type have apparently first been
considered in \cite{fro2}. For the special case of $Z(N)$ order disorder duality they
date back at least to \cite{swi}.\\
2. By this result the field extension of Definition \ref{def3} in conjunction with
Theorem \ref{thm:action} may be considered a local version of the
construction of the double. (If we had used the $U_R^\2O(g)$ we would
have ended up with $R^{-1}$ which would do just as well.)\\
3. If the net $\2O\mapsto\2F(\2O)$ is fermionic an additional sign $\pm$
appears on the right hand side of (\ref{commrel3}) depending on the Bose/Fermi
nature of the fields. Using the bosonization prescription of the next section
this sign can be eliminated.

We now turn to a discussion of the localized endomorphisms of the observable algebra
$\2A$ which are implemented by the charged fields in $\hat{\2F}$ as in \cite{dr1}.
Let $\psi_i,\ i=1,\ldots, d_\psi$ be a multiplet of isometries in $\hat{\2F}(\2O)$
transforming according to the irreducible representation $r$ of $D(G)$. Then the map 
\be \rho(\cdot) = \sum_{i=1}^{d_\psi} \psi_i \cdot \psi_i^* \label{rho}\ee
defines a unital *-endomorphism of $\hat{\2F}$. The relative locality of $\2A$ 
and $\hat{\2F}$ implies the restriction of $\rho$ to $\2A$ to be localized 
in $\2O$ in the sense that $\rho(A)=A\ \forall A\in\2A(\2O')$. 
Furthermore, $\rho$ maps $\2A(\2O_1)$ into itself if $\2O_1\supset\2O$ as 
follows from the $D(G)$-invariance of $\rho(x)$ for $x\in\2A$. 
(The conventional argument using duality would allow us only to conclude 
$\rho(\2A(\2O_1))\subset\hat{\2A}(\2O_1)$.)
\bprop In restriction to $\2A(\2O_1),\ \2O_1\supset\2O$ the 
endomorphism $\rho$ is irreducible. \label{irr}\eprop
\prf The proof is omitted as it is identical to the proof of \cite[Prop. 6.9]{lo1} 
where compact groups are considered. \qed\\
\rems 1. In application to the net $\hat{\2A}$ the endomorphisms $\rho$ are localized
only in wedge regions, i.e.\ they are of solitonic character.\\
2. Due to the spontaneous breakdown of the quantum symmetry the endomorphisms 
$\rho$ which arise from non-group representations of $D(G)$ should not be considered
as true superselection sectors of the net $\2A\restr\2H_0$. This would be justified 
if the symmetry were unbroken. Nevertheless, one can analyze their statistics, as
will be done in the rest of this subsection.

Whereas the endomorphisms $\rho$ defined above need not be invertible one can always
find left inverses \cite{dhr1} $\phi$ such that $\phi\circ\rho=id$. For $\rho$ as 
defined by (\ref{rho}) the left inverse is easily verified to be given by
\be \phi_\rho=\frac{1}{d_\psi}\sum_{i=1}^{d_\rho} \psi_i^* \cdot \psi_i 
  \label{leftinv}. \ee

In order to study the statistics of endomorphisms one introduces \cite{dhr3,frs1} 
the statistics operators
\be \ve(\rho_1,\rho_2)=U_2^*\,\rho_1(U_2)\in(\rho_1\rho_2,\rho_2\rho_1), \ee
where $U_2$ is a charge transporter intertwining $\rho_2$ and $\tilde{\rho}_2$, 
the latter being localized in the left spacelike complement of the 
localization region of $\rho_1$. Such an intertwiner is given by
$U_2=\sum_i \tilde{\psi}^2_i\psi_i^{2*}\in\2F^G=\2A$, where 
$\tilde{\psi}_i$ is a multiplet in $\hat{\2F}(\tilde{\2O}),\ \tilde{\2O}<\2O_1$
transforming according the same representation of $D(G)$ as $\psi_i$, such that $U_2$
is $D(G)$-invariant and thus in $\2A$.
\blemma Let $\psi^1_i\in\hat{\2F}(\2O_1), i=1,\ldots, d_1$ and 
$\psi^2_j, j=1,\ldots, d_2$ be
$D(G)$-multiplets corresponding to the representations $D^1, D^2$ and let 
$\rho_1, \rho_2$ be the associated endomorphisms.
Then the statistics operator is given by
\be \ve(\rho_1,\rho_2)=\sum_{ijkl} \psi_i^2\,\psi_l^1\,\psi_j^{2*}\,\psi_k^{1*}\ 
  (D_{lk}^1\otimes D_{ij}^2)(R) .\label{statop}\ee
The statistics parameter \cite{dhr1} for the morphism $\rho$ which is implemented by 
the irreducible $D(G)$-tensor $\psi_i, i=1,\ldots, d_\psi$ is 
\be \lambda_\rho = \frac{\omega_\rho}{d_\rho},\label{statparam}\ee
with $d_\rho=d_\psi$ and $D_{lj}(X)=\delta_{lj}\,\omega_\rho$ where 
$X=\sum_g V_g\,U_g$ is a unitary element in the center of $D(G)$.
\elemma
\prf With $U_2=\sum_i \tilde{\psi}^2_i\psi_i^{2*}$ we have
\be \ve(\rho_1,\rho_2)=\sum_{ijk} \psi^2_i\tilde{\psi}_i^{2*} \ \psi_j^1 \,
   \tilde{\psi}^2_k\psi_k^{2*} \, \psi_j^{1*}. \ee
Then (\ref{statop}) follows by an application of (\ref{commrel3}) to $\psi_j^1$
and $\tilde{\psi}^2_k$ and appealing to the orthogonality relation 
$\tilde{\psi}^{2*}_i\tilde{\psi}^2_j=\delta_{ij}\11$.
With the identification $\psi^1=\psi^2=\psi$ in (\ref{statop}), (\ref{leftinv}) and using
once more the orthogonality relation we compute the statistics parameter as follows:
\be \lambda_\rho\,\11=\phi_\rho(\ve_{\rho,\rho})=
   \frac{1}{d_\psi}\sum_{ijl}\psi_l\,\psi_j^*\ (D_{li}\otimes D_{ij})(R) =
   \frac{1}{d_\psi}\sum_{jl}\psi_l\,\psi_j^*\ M_{lj}, \ee
where 
\be M_{lj}=\sum_i (D_{li}\otimes D_{ij})(R) =D_{lj}(\sum_g V_g\,U_g). \ee
An easy calculation shows that $X=\sum_g V_g\,U_g$ is a unitary element in the
center of $D(G)$ such that it is represented by a phase times the unit matrix in the 
irreducible representation $D$: $M_{lj}=\delta_{lj}\,\omega,\ \omega\in S^1$. \qed\\
\rems 1. The statistical dimension of the sector $\rho$, defined as 
$d_\rho=|\lambda_\rho|^{-1}$ coincides with the dimension of the corresponding 
representation of the quantum double.
This was to be expected and is in accord with the fact \cite{lo2} that the action 
of finite dimensional Hopf algebras cannot give rise to non-integer dimensions.\\
2. Recalling Lemma \ref{tensor} we see that in restriction to a field operator
in a multiplet transforming according to the irreducible representation
$r$ the action of $\gamma_X$ amounts to multiplication by $\omega_r$.
The unitary $X\in D(G)$ may thus be interpreted as the quantum double analogue of 
the group element $k$ which distinguishes between bosons and fermions. This is 
reminiscent of the notion of ribbon elements in the framework of quantum groups, see 
Appendix A. In fact, the operator $X$ defined above is just the inverse of Drinfel'd's 
$u=\sum_g V_g\,U_{g^{-1}}$ which itself is a ribbon element due to $S(u)=u$.\\
3. Appealing to the representation theory of $D(G)$ 
as expounded in \cite{dpr} it is easy to compute the phase $\omega_r$ for the 
representation $r=(c,\pi)$.
It is given by the scalar to which $g\in c$, obviously being contained in the center 
of the normalizer $N_g$, is mapped by the irreducible representation $\pi$ of $N_g$.
As an immediate consequence \cite{dvvv} $\omega_r$ is an $n$-th root of unity where 
$n$ is the order of $g$. 

We now turn to the calculation of the monodromy operator
\be \ve_M(\rho_1,\rho_2) = \ve(\rho_1,\rho_2)\,\ve(\rho_2,\rho_1) \label{monodr}\ee
which measures the deviation from permutation group statistics and of the
{\it statistics characters} \cite{khr}
\be Y_{ij}\,\11=d_i d_j\,\phi_i(\ve_M(\rho_i,\rho_j)^*) .\label{statchar}\ee
In the latter expression $\rho_i, \rho_j$ are irreducible morphisms such that
the right hand side is a C-number since 
$\phi_i(\ve_M(\rho_i,\rho_j)^*)$ is a self\/intertwiner of $\rho_j$ and due to the 
irreducibility of the latter, cf.\ Proposition \ref{irr}.)
We thus obtain a square matrix of complex numbers indexed by the superselection sectors, 
i.e.\ in our case the irreducible representations of the quantum double $D(G)$.
\bprop In terms of the fields the monodromy operator is given by
\be \ve_M(\rho_1,\rho_2)=\sum_{ijkl} \psi_i^2\,\psi_j^1\,\psi_k^{1*}\,\psi_l^{2*}\ 
  (D^1_{jk}\otimes D^2_{il})(I), \label{monodr1}\ee
where
\be I=R\,\sigma(R). \label{defI}\ee
The statistics characters are given by
\be Y_{ij}=(tr_i\otimes tr_j)\circ(D^i\otimes D^j)(I^*) .\label{defY}\ee
\eprop
\prf Inserting the statistics operators according to (\ref{statop}) and using twice the 
orthogonality relation we obtain
\be \ve_M(\rho_1,\rho_2)=\sum_{\ba{c} ijkl\\ k'l'\ea} \psi_i^2\,\psi_j^1\,\psi_{k'}^{1*}
  \,\psi_{l'}^{2*}\ (D^1_{jl}\otimes D^2_{ik})(R)\,(D^2_{kl'}\otimes D^1_{lk'})(R).
\label{monodr2}\ee
The numerical factor to the right (including the summations over $k, l$) can be
simplified to
\be \sum_{g,h\in G} D^1_{jk'}(V_g\,U_h)\,D^2_{il'}(U_g\,V_h)=
  (D^1_{jk'}\otimes D^2_{il'})(I) .\ee
Omitting the primes on $k', l'$ we obtain (\ref{monodr1}). The formula (\ref{defY})
follows in analogy to the computation of $\lambda_\rho$ from (\ref{monodr1}), 
(\ref{leftinv}) and $Y_{ij}\propto\11$. \qed\\
\rem $I=R\,\sigma(R)$ can be considered as the quantum group version of the monodromy 
operator.

In \cite{ac} it was shown that $Y$ is invertible, in fact $\frac{1}{|G|}Y$ is unitary.
In conjunction with the known facts concerning the representation theory one concludes
\cite{ac} that the quantum double
$D(G)$ is a {\it modular Hopf algebra} in the sense of \cite{rt}. We are now in a
position to complete our demonstration of the complete parallelism between quantum 
group theory and quantum field theory (which we claim only for the quantum double 
situation at hand!).
What remains to be discussed is the Verlinde algebra structure \cite{v} behind the
fusion of representations of the double and the associated endomorphisms of 
$\hat{\2F}$, respectively.
The fusion rules are said to be diagonalized by a unitary matrix $S$ if
\be N_{ij}^k= \sum_m \frac{S_{im}\,S_{jm}\,S^*_{km}}{S_{0m}}. \label{fusion}\ee
(For a comprehensive survey of fusion structures see \cite{fuchs}.)
One speaks of a Verlinde algebra if, in addition, $S$ is symmetric, there is a
diagonal matrix $T$ of phases satisfying $TC=CT=T$ 
($C_{ij}=\delta_{i\bar{\jmath}}$ is the charge conjugation matrix) and $S$ and $T$
constitute a representation of $SL(2,\7Z)$ (in general not of 
$PSL(2,\7Z)=SL(2,\7Z)/\7Z_2$), i.e.
\be S^2=(ST)^3=C .\label{verlinde}\ee
On the one hand the representation categories of modular Hopf algebras are known 
\cite{rt} to be modular, i.e.\ to satisfy (\ref{fusion}) and (\ref{verlinde}), where the
phases in $T$ are given by the values of the ribbon element $X$ in the irreducible
representations.

On the other hand this structure has been shown \cite{khr} to arise from the 
superselection structure of {\it every} rational quantum field theory in $1+1$ 
dimensions. In this framework the phases in $T$ are given by the phases of the
statistics parameters (\ref{statparam}), whereas the matrix $S$ arises from the
statistics characters
\be T=\left(\frac{\sigma}{|\sigma|}\right)^{1/3}\mbox{Diag}(\omega_i), \ \ 
   S=|\sigma|^{-1}\, Y .\ee
For nondegenerate theories the number $\sigma=\sum_i \omega_i^{-1}\,d_i^2$ satisfies 
$|\sigma|^2=\sum_i d_i^2$. Using the result \cite{ac} $\sigma=|G|$ this condition is 
seen to be fulfilled, for the semisimplicity of $D(G)$ gives
$\sum_i d_i^2=\mbox{dim}(D(G))=|G|^2$.

We thus observe, for the orbifold theories under study, a perfect parallelism between 
the general superselection theory \cite{khr} for quantum field theories in low 
dimensions and the representation theory of the quantum double \cite{dpr}. This 
parallelism
extends beyond the Verlinde structure. One observes, e.g., that the equations (2.4.2)
of \cite{dpr} and (2.30) in \cite{khr}, both stating that the monodromy operator
is diagonalized by certain intertwining operators, are identical although
derived in apparently unrelated frameworks.

\sectreset{Bosonization}
In this section we will show how the methods expounded in the preceding sections
can be used to obtain an understanding of the Bose/Fermi correspondence in 1+1
dimensions in the framework of local quantum theory. This is so say, we 
will show how one can pass from a fermionic net of algebras with twisted duality to 
a bosonic net satisfying
Haag duality {\it on the same Hilbert space}, and vice versa. Our method amounts
to a continuum version of the Jordan-Wigner transformation and is reminiscent of
Araki's approach to the XY-model \cite{araki2}. 

Our starting point is as defined in the introduction, i.e.\ a net of field algebras
with fermionic commutation relations (\ref{commrel}) and 
twisted duality (\ref{twduality}) augmented by the
split property for wedge regions introduced in Section 2. As before there exists a 
selfadjoint unitary operator $V$ distinguishing between even and odd operators.
For the present investigations, however, the existence of further inner symmetries is 
ignored as they are irrelevant for the spacelike commutation relations.
Therefore we now repeat the field extension of Section 3 replacing the group $G$
by the subgroup $\7Z_2=\{e,k\}$. This amounts to simply extending the local algebras
by the disorder operator associated with the only nontrivial group element $k$
\be \hat\2F(\2O)=\2F(\2O) \vee \{ V^\2O \} ,\ee
where $V^\2O=U^\2O_L(k)$. Again, the assignment $\2O\mapsto\hat{\2F}(\2O)$
is isotonous, i.e.\ a net. This is of course the simplest instance of the 
situation discussed at the beginning
of Section 4 where it was explained that there is an action of the dual group 
$\hat{G}$ on the extended net. We thus have an action of $\7Z_2\times\7Z_2$ on
the quasilocal algebra $\hat{\2F}$ generated by $\alpha=Ad\, V$ and $\beta$
\bea \alpha(F+GV^\2O) &=& F_+-F_-+(G_+-G_-)V^\2O, \\
  \beta (F+GV^\2O) &=& F-GV^\2O \eea
where $F, G\in\2F$.
We now define $\tilde{\2F}(\2O)$ as the fixpoint algebra under the {\it diagonal}
action $\alpha\circ\beta=\beta\circ\alpha$:
\be \tilde{\2F}(\2O)=\{ x\in\hat{\2F}(\2O)\mid x=\alpha\circ\beta(x) \} .\ee
Obviously $\tilde{\2F}(\2O)$ can be represented as the following sum:
\be \tilde{\2F}(\2O)=\2F(\2O)_+  +  \2F(\2O)_- \, V^\2O .\label{ftilde}\ee
It is instructive to compare $\tilde{\2F}(\2O)$ with the twisted algebra
\be \2F(\2O)^t=\2F(\2O)_+  +  \2F(\2O)_-\, V ,\ee
the only difference being that in the former expression $V^\2O$ appears instead
of $V$. This reflects just the difference between Jordan-Wigner and Klein
transformations.
It is well known that the net $\2F^t$ is local relative to $\2F$. That
the former cannot be local itself, however, follows clearly from the fact that it
is unitarily equivalent to the latter by $\2F(\2O)^t=Z\2F(\2O) Z^*$.
\blemma Let $W_L$ and $W_R$ be left and right wedges, respectively.
Then the wedge algebras of $\tilde{\2F}$ are given by
\bea \tilde{\2F}(W_L)&=&\2F(W_L),\\ \tilde{\2F}(W_R)&=&\2F(W_R)^t .\eea
Wedge duality holds for the net $\tilde{\2F}$.
\label{o-jwt}\elemma
\prf $V^\2O$ is contained in $\2F(W_L)_+$ for any $\2O\subset W_L$. Thus,
$\2F(W_L)_-\, V^\2O=\2F(W_L)_-$, whence the first identity. Similarly we have
$V^\2O_R\in\2F(W_R)_+$ for $\2O\in W_R$, from which we obtain
$\2F(W_L)_-\, V^\2O=\2F(W_L)_- \, V$. Wedge duality for $\tilde{\2F}$ now
follows immediately from twisted duality for $\2F$.\qed
\bprop The net $\2O\mapsto\tilde{\2F}(\2O)$ is local.
\eprop
\prf Let $\2O_1, \2O_2$ be mutually spacelike double cones. We may assume
$\2O_1 < \2O_2$ such that $W^{\2O_1}_L$ and $W^{\2O_2}_R$ are mutually spacelike.
The commutativity of $\tilde{\2F}(\2O_1)$ and $\tilde{\2F}(\2O_1)$ follows from the 
preceding lemma and twisted locality for $\2F$ since $\2O_1\subset W^{\2O_1}_L$ and 
$\2O_2\subset W^{\2O_2}_R$. \qed\\
\rem A more intuitive proof goes as follows. Let $F_i\in\2F(\2O_i)_-, i=1,2$.
Then commuting $F_1\,V^{\2O_1}$ through $F_2\,V^{\2O_2}$ 
gives exactly two factors of $-1$. The first arises from $F_1 F_2=-F_2 F_1$
and the other from $V^{\2O_2} F_1=-F_1 V^{\2O_2}$, whereas 
$V^{\2O_1} F_2=F_2 V^{\2O_1}$.
\bprop The net $\tilde{\2F}$ fulfills Haag duality for double cones.
\eprop
\prf We have to prove 
$\tilde{\2F}(\2O)=\tilde{\2F}(W^\2O_L)\wedge\tilde{\2F}(W^\2O_L)$.
Using the Lemma the right hand side is seen to equal $\2F(W^\2O_L)\wedge\2F(W^\2O_R)^t$
which by (\ref{newisom2}) is unitarily equivalent to 
$\2F(W^\2O_R)^t\otimes\2F(W^\2O_L)$. On the other hand (\ref{newisom3}) leads to
\bea \lefteqn{\tilde{\2F}(\2O) = \2F(\2O)_+ + \2F(\2O)_-\,V^\2O } \nn\\ 
   && \cong \ \2F(W^\2O_R)_+ \otimes \2F(W^\2O_L)_+ \  + \
   \2F(W^\2O_R)_-\,V \otimes \2F(W^\2O_L)_- \nn\\
   &&  +  \  [ \2F(W^\2O_R)_- \otimes \2F(W^\2O_L)_+ \  + \
   \2F(W^\2O_R)_+\,V \otimes \2F(W^\2O_L)_- ] \, V\otimes \11  \nn\\
   && = \2F(W^\2O_R)^t \otimes \2F(W^\2O_L) \eea
which completes the proof.  \qed

It is obvious that the net $\tilde{\2F}$ is \poinc\ covariant with respect to 
the original representation of $\2P$. Finally, the group $G$ acts on $\tilde{\2F}$
via the adjoint representation $g\mapsto Ad\,U(g)$. In particular $Ad\,U(k)=Ad\,V$
acts trivially on the first summand of the decomposition (\ref{ftilde}) and 
by multiplication with $-1$ on the second, i.e.\ the bosonized theory carries
an action of $\7Z_2$ in a natural way.

It should be clear that the same construction can be used to obtain a twisted dual
fermionic net from a Haag dual bosonic net with a $\7Z_2$ symmetry. It is not
entirely trivial that these operations performed twice lead back to the net one
started with, as the operators $V^\2O$ constructed with the original and the
bosonized net might differ. That this is not the case, however, can be derived from 
Lemma \ref{o-jwt}, the easy argument is left to the reader.

\sectreset{Conclusions and Outlook}
In this final section we summarize our results and relate them to some of those in the 
literature. Starting from a local \qft\ in $1+1$ dimensions with an {\it unbroken} 
group symmetry we have discussed disorder operators which implement a global
symmetry on some wedge region and commute with the operators localized in the spacelike 
complement of a somewhat larger wedge. Whereas disorder operators are only 
localized in wedge regions, they can in a natural way be associated to the bounded 
region where the interpolation between the global group action and the trivial action 
takes place. Extending the local algebras $\2F(\2O)$ of the original theory by the 
disorder operators corresponding to the double cone $\2O$ gives rise to a nonlocal net 
$\hat{\2F}$ which is uniquely defined. We have shown that for every \qft\ fulfilling 
a sufficiently strong version of the split property disorder operators exist and can 
be chosen such as to transform nicely under the global group action. As a consequence, 
the extended theory supports an action
of the quantum double $D(G)$ which, however, is spontaneously broken in the sense that 
only the subalgebra $\7CG$ is implemented by operators on the Hilbert space. 
Nevertheless, all other aspects of the quantum symmetry, like R-matrix commutation 
relations and the Verlinde algebra, show up and correspond nicely to the structures 
expected due to the general analysis \cite{frs1,khr}. The spontaneous breakdown of the
quantum symmetry is in accord with the findings of \cite{km} where it was argued 
(in the case of a cyclic group $Z(N)$) that the vacuum expectation values of order 
and disorder variables can vanish jointly, as they must in the case of unbroken quantum
symmetry, only if there is no mass gap. Massless theories are, however, ruled out by
the postulate of the split property for wedges upon which our analysis hinges.

The fact that in the situation studied in this paper `one half' of the quantum double 
symmetry is spontaneously broken hints at an alternative construction which we describe
briefly. Given a local net of $C^*$-algebras with group symmetry there may of course 
be vacuum states which are {\it not} gauge invariant. Let us assume that $\omega_e$ 
is such that $\omega_g=\omega_e\circ\alpha_g\ne\omega_e\ \forall g\ne e$, i.e.\ the 
symmetry is completely broken. One may now consider the reducible representation 
$\oplus_g \pi_g$ of $\2F$ on the Hilbert space $\hat{\2H}=L^2(G,\2H)$ where $\pi_g$ is
the GNS-representation corresponding to a soliton state which connects the
vacua $\omega_e$ and $\omega_g$. The existence of such states follows from the 
same set of assumptions as was used in the present investigation \cite{schl}.
Again, one can construct operators $U^\2O(g)$ enjoying similar algebraic properties
as the disorder operators appearing in this paper. Their interpretation is different, 
however, in that they are true soliton operators intertwining the vacuum representation 
and the soliton sectors. Extending the local algebras according
to (\ref{def3}) gives rise to a field net $\hat{\2F}$ which acts irreducibly on
$\hat{\2H}$. The details of the construction outlined above, which is complementary 
in many ways to the one studied in the present work, will be given in a forthcoming 
publication. In the solitonic variant there is also an action of the quantum double
$D(G)$ where the action of $C(G)$ is implemented in the obvious way, whereas the 
group symmetry is spontaneously broken.

Although the split property for wedges should be satisfied by reasonable massive \qfts\ 
it definitely excludes conformally invariant models, which via \cite{dvvv,dpr} provided
part of the motivation for the present investigation. Concerning this somewhat 
disturbing point we confine ourselves to the following remarks.
It is well known that \qfts\ in $1+1$ dimensions, like the $\2P(\phi)_2$ models,
possess a unique symmetric vacuum for some range of the parameters whereas 
spontaneous symmetry breakdown and vacuum degeneracy occur for other choices.
The construction sketched above shows that the algebraic structure of order/disorder
duality is the same in both massive regimes.
It is furthermore known that the $\2P(\phi)_2$ theory with interaction 
$\lambda\phi^4-\delta\phi^2$ possesses a critical point at the
interface between the symmetric and broken phases. Unfortunately, little is rigorously
known about the possible conformal invariance of the theory at this point. The
case of conformal invariance is quite different anyway, for Haag duality of the net 
$\2F$ and of the fixpoint net $\2A$ are compatible in contrast to the massive case.

In the framework of lattice models things are easier as the local degrees of freedom
are amenable to more direct manipulation. The authors of \cite{szlvec} considered
a class of models where the disorder as well as the order operators were explicitly 
defined by specifying their action on the Hilbert space associated to a finite region. 
They then had to assume the existence of a vacuum state which is invariant under the 
action of the quantum double. In his approach \cite{araki2} to the XY-model Araki 
similarly defines an automorphism of the algebra of order variables which is 
localized in a halfspace and then constructs the crossed product. In the continuum
solitonic automorphisms can be defined for some models \cite{fro1}, but for a model
independent analysis there seems to be no alternative to our abstract approach.

As to the interpretation of the structures found in the present work and outlined
above, we have already remarked that they may be considered as a local version of
the construction of the quantum double. The quantum double was invented by
Drinfel'd as a means to obtain quasitriangular Hopf algebras, and in \cite{reshet}
it was shown to be `factorizable', see Appendix A. Furthermore, every finite 
dimensional factorizable Hopf algebra can be obtained as a quotient of a quantum 
double by a twosided ideal. One may therefore expect that quantum doubles will play 
an important role in an extension of the constructions in \cite{dr2} to low dimensional
theories.\\ \\ \noindent
{\it Acknowledgments.} I am greatly indebted to K.-H.~Rehren for his stimulating
interest, many helpful discussions and countless critical readings of the evolving 
manuscript. Special thanks are due to D.~Buchholz for the proof of Proposition 
\ref{outer}.

\appendix
\sectreset{Quantum Groups and Quantum Doubles}
A Hopf algebra is an algebra $H$ which at the same time is a coalgebra, i.e.\ there are
homomorphisms $\Delta: H\to H\otimes H$ and $\ve: H\to \7C$ satisfying
\be (\Delta\otimes\mbox{id})\circ\Delta=(\mbox{id}\otimes\Delta)\circ\Delta ,\ee
\be (\ve\otimes\mbox{id})\circ\Delta=(\mbox{id}\otimes\ve)\circ\Delta=\mbox{id} ,
   \label{counit0}\ee
with the usual identification $H\otimes\7C=\7C\otimes H=H$. Furthermore, there is an 
antipode, i.e.\ an antihomomorphism $S: H\to H$ for which
\be m\circ(S\otimes\mbox{id})\circ\Delta=m\circ(\mbox{id}\otimes S)\circ\Delta=
   \ve(\cdot) \11 ,\ee
where $m:\,H\otimes H\to H$ is the multiplication map of the algebra.\\
\rem By (\ref{counit0}) the counit, which is simply a one dimensional representation, 
is the `neutral element' with respect to the comultiplication.

For the quantum double $D(G)$ defined in Definition \ref{double1} these maps are given
by
\bea \ve(V(g)U(h)) &=& \delta_{g,e} ,\label{counit}\\
 \Delta(V(g)U(h)) &=& \sum_k V(hk)U(h)\otimes V(k^{-1})U(h) ,\label{coprod}\\
 S(V(g)U(h)) &=& V(h^{-1}g^{-1}h)U(h^{-1}) \label{antip}\eea
on the basis $\{V(g)U(h)\,g,h\in G\}$ and extended to $D(G)$ by linearity.

A Hopf algebra $H$ is {\it quasitriangular}, or simply a quantum group, if there is an
element $R\in H\otimes H$ satisfying
\be \Delta'(\cdot)=R\,\Delta(\cdot)\,R^{-1} ,\ee
where $\Delta'=\sigma\circ\Delta$ with $\sigma(a\otimes b)=b\otimes a$ and
\bea (\Delta\otimes\mbox{id})(R) &=& R_{13}\,R_{23} ,\label{r1}\\
 (\mbox{id}\otimes\Delta)(R) &=& R_{13}\,R_{12} .\label{r2}\eea
Here $R_{12}=R\otimes\11, R_{23}=\11\otimes R$ and 
$R_{13}=(id\otimes\sigma)(R\otimes\11)$. As a consequence, $R$ satisfies the 
Yang-Baxter equation
\be R_{12}\,R_{13}\,R_{23} = R_{23}\,R_{13}\,R_{12} .\ee
It is easy to verify that the R-matrix (\ref{rmatrix}) satisfies these requirements.\\
\rem As shown by Drinfel'd, for quantum groups the square of the antipode is inner,
i.e.\ $S^2(a)=uau^{-1}$ where $u$ is given by
$u=m\circ(S\otimes\mbox{id})\circ\sigma(R)$. The operator $u$ satisfies
$\ve(u)=1,\ \Delta(u)=(\sigma(R)R)^{-1}\, (u\otimes u)=(u\otimes u)(\sigma(R)R)^{-1}$.
For quantum doubles of finite groups the antipode is even involutive 
($S^2=\mbox{id}$, equivalently 
$u$ is central). This holds for all finite dimensional Hopf-\1*-algebras, 
whether quantum groups or not.

A quantum group is called {\it factorizable} \cite{reshet} if the map $H^*\to H$ given
by $H^*\ni x\mapsto\langle x\otimes id,I\rangle$ is nondegenerate, where $I$ is as in
(\ref{defI}). Quantum doubles are automatically factorizable.

A quasitriangular Hopf algebra possessing a (non unique) central element $v$ satisfying
the conditions
\be v^2=u\,S(u),\ \ve(v)=1,\, S(v)=v ,\ee
\be \Delta(v)=(\sigma(R)R)^{-1}\, (v\otimes v) ,\ee
where $u$ is the operator defined in the above remark,
is called a {\it ribbon Hopf algebra} \cite{rt}. 

Finally, {\it modular Hopf algebras} are defined by some restrictions on their 
representation structure, the most important of which is the nondegeneracy of the
matrix $Y$ defined in (\ref{defY}). Obviously, the conditions of factorizability
and modularity are strongly related. 

\sectreset{Generalization to Continuous Groups}
In this appendix we will generalize our considerations on quantum double actions to
arbitrary locally compact groups (the quantum field theoretic framework gives rise
only to compact groups.)
In Section 4 we identified von Neumann algebras acted upon by the double 
$D(G)$ of a finite group with von Neumann algebras which are simultaneously graded 
by the group and automorphically acted upon by the latter, satisfying in addition 
the relation (\ref{coupl}). The concept of group grading, however, loses its meaning
for continuous groups. This problem is solved by appealing to the well known fact 
(see e.g.\ the introduction to \cite{landst}) that an algebra $A$ (von Neumann or 
unital $C^*$) graded by a finite group $G$ is the same as an algebra with a coaction 
of the group. A coaction is a homomorphism $\delta$ from $A$ into $A\otimes\7CG$ 
satisfying
\be (\delta\otimes id)\circ\delta=(id \otimes \delta_G)\circ\delta ,\label{coact-0}\ee
where $\delta_G: \7CG\rightarrow\7CG\otimes\7CG$ is the coproduct given by 
$g\mapsto g\otimes g$. The correspondence between these notions is as follows.
Given a $G$-graded algebra $A=\oplus_g A_g,\ A_g A_h\subset A_{gh}$
and defining $\delta(x)=x\otimes g$ for $x\in A_g$ one obtains a coaction. The
converse is also true. The relation $\alpha_g(A_h)=A_{ghg^{-1}}$ between the
group action and the grading obviously translates to
\be \delta\circ\alpha_g=(\alpha_g\otimes Ad\,g)\circ\delta .\label{coact}\ee

The concept of coaction extends to continuous groups, 
where the group algebra $\7CG$ is replaced by the von Neumann algebra $\2L(G)$ 
(here we will treat only quantum double actions on von Neumann algebras) of the left 
regular representation which is generated
by the operators $(\lambda(g)\xi)(h)=\xi(g^{-1}h)$ on the Hilbert space $L^2(G)$.

In the next step we give a precise definition of the double of continuous group.
To this purpose we have to put a topology on the crossed product 
of some algebra of functions on the group by the adjoint action of the latter.
There are many ways of doing this, as is generally the 
case with infinite dimensional vector spaces. For compact {\it Lie} groups two 
different constructions, one of which appears to generalize to arbitrary compact 
groups, have been given in in \cite{bonn2}. The most important virtue of this work 
is that the topological Hopf algebras obtained there are reflexive as topological 
vector spaces, making the duality between $D(G)$ and $D(G)^*$ very explicit. From 
the technical point of view, however, the Fr\'{e}chet topologies on which this
approach relies are not very convenient. 

In the following we will define the quantum double in the framework of Kac algebras
\cite{en-schw,en-va}. The latter has been invented as a generalization of locally 
compact groups which is closed under duality. As the $C^*$ and von Neumann versions 
of Kac algebras have been proved \cite{en-va} equivalent (generalizing the equivalence
between locally compact groups and measurable groups) it is just a matter of
convenience which formulation we use. We therefore consider first the von Neumann
version which is technically easier. 

We start with the von Neumann algebra $M=L^\infty(G)$ of essentially bounded 
measurable functions acting on the Hilbert space $H=L^2(G)$ by pointwise 
multiplication.
With the coproduct $\Gamma(f)(g,h)=f(gh)$ and the involution $\kappa(f)(g)=f(g^{-1})$
it is a coinvolutive Hopf von Neumann algebra. This means $\Gamma$ is a coassociative
isomorphism of $M$ into $M\otimes M$, $\kappa$ is an anti-automorphism
(complex linear, antimultiplicative and $\kappa(x^*)=\kappa(x)^*$) and 
$\Gamma\circ\kappa=\sigma\circ(\kappa\otimes\kappa)\circ\Gamma$ holds where 
$\sigma$ is the flip.
The weight $\vp$, defined on $M_+$ by $\vp(f)=\int_G dg\, f(g)$, 
is normal, faithful, semifinite (n.f.s.) and fulfills
\begin{enumerate}
\item For all $x\in M_+$ one has $(\imath\otimes\vp)\Gamma(x)=\vp(x)\11$.
\item For all $x,y\in\mathfrak{n}_\vp$ one has
$(\imath\otimes\vp)((\11\otimes y^*)\Gamma(x))=\kappa((\imath\otimes\vp)(\Gamma(y^*)
(\11\otimes x)))$.
\item $\kappa\circ\sigma^\vp_t=\sigma^\vp_{-t}\circ\kappa\ \forall t\in\7R$.
\end{enumerate}
This makes $(M,\Gamma,\kappa,\vp)$ a Kac algebra in the sense of \cite{en-schw},
well known as $KA(G)$. The dual Kac algebra \cite{en-schw} 
of $KA(G)$ is $KS(G)=(\2L(G),\hat{\Gamma},\hat{\kappa},\hat{\vp})$, the
von Neumann algebra of the left regular representation equipped with the coproduct 
$\hat{\Gamma}(\lambda(g))=\lambda(g)\otimes\lambda(g)$, the coinvolution
$\hat{\kappa}(\lambda(g))=\lambda(g^{-1})$ and the weight $\hat{\vp}$ 
which we do not bother to state (see e.g.\ \cite{haage}).

Defining now an action of $G$ on $M$ by
the automorphisms $\alpha_g(f)(h)=f(g^{-1}hg)$ it is trivial to check weak
continuity with respect to $g$. Furthermore, $\alpha_g$ is unitarily implemented by
$u_g=\lambda(g)\rho(g)$, where $(\rho(g)\xi)(h)=\Delta(g)^{1/2}\xi(hg)$ is the right 
regular representation.
We can thus consider the crossed product (in the usual
von Neumann algebraic sense \cite{takes}) $\tilde{M}=M\rtimes_\alpha G$ on 
$H\otimes L^2(G)\ (=L^2(G)\otimes L^2(G))$, generated by $\pi(M)$ and
$\lambda_1(g)=\11_M\otimes\lambda(g),\ g\in G$.
\bprop There are mappings $\tilde{\Gamma},\tilde{\kappa},\tilde{\vp}$ on
$\tilde{M}$ such that the quadruple 
$(\tilde{M},\tilde{\Gamma},\tilde{\kappa},\tilde{\vp})$ is
a Kac algebra, which we call the quantum double $\2D(G)$. On the subalgebras
$\pi(M)$ and $\lambda_1(G)''=\11_M\otimes\2L(G)$ the coproduct and the coinvolution 
act according to
\bea \tilde{\Gamma}(\pi(x))=(\pi\otimes\pi)(\Gamma(x)) ,
  & \tilde{\kappa}(\pi(x))=\pi(\kappa(x)), & x\in M,\label{double-1}\\
  \tilde{\Gamma}(\lambda_1(g))=\lambda_1(g)\otimes\lambda_1(g) , 
  & \tilde{\kappa}(\lambda_1(g))=\lambda_1(g^{-1}), & g\in G.\label{double-2}\eea
The Haar weight $\tilde{\vp}$ is given by the dual weight \cite{haage}
\be \tilde{\vp}=\vp\circ\pi^{-1}\circ(\imath_{\tilde{M}}\otimes\hat{\vp})
   (\tilde{\delta}(x)) ,\ee
where $\tilde{\delta}$ is the dual coaction from $\tilde{M}$ to 
$\tilde{M}\otimes\2L(G)$ which acts according to
\bea \tilde{\delta}(\pi(x)) &=& \pi(x)\otimes \11_{\2L(G)}, \ \ \ x\in M \\
   \tilde{\delta}(\lambda_1(g)) &=& \lambda_1(g)\otimes\lambda(g),\ \ \ g\in G .\eea
\eprop
\prf The automorphisms $\alpha_g$ of $M$ are easily shown to satisfy 
$\Gamma\circ\alpha_g=(\alpha_g\otimes\alpha_g)\circ\Gamma$ and 
$\kappa\circ\alpha_g=\alpha_g\circ\kappa$. (The first identity is just 
$g^{-1}(hk)g=(g^{-1}hg)(g^{-1}kg)$, the second $(g^{-1}hg)^{-1}=g^{-1}h^{-1}g$.)
Thus $\alpha: G\rightarrow Aut\,M$ constitutes an action of $G$ on the Kac algebra
$(M,\Gamma,\kappa,\vp)$ in the sense of \cite{cann}. We can now apply
\cite[Th\'{e}or\`{e}me 1]{cann} to conclude that there exist a coproduct, a 
coinvolution and a Haar weight on $\tilde{M}$ such that the axioms of a Kac algebra
are satisfied. The equations (\ref{double-1},\ref{double-2}) are restatements of 
\cite[Propositions 3.1, 3.3]{cann} whereas the Haar weight is as in 
\cite[D\'{e}finitions 1.9]{cann}. \qed
\bprop The dual Kac algebra of the quantum double is
$\widehat{\2D(G)}=(\2L(G)\otimes L^\infty(G),\hat{\tilde{\Gamma}},
\hat{\tilde{\kappa}},\hat{\vp}\otimes\vp)$. The coproduct and the counit are
\bea \hat{\tilde{\Gamma}} (x) &=& R\,(\11\otimes\sigma\otimes\11)\,(\hat{\Gamma}
   \otimes\Gamma)(x)\,(\11\otimes\sigma\otimes\11)\,R^* \label{gammahat}\\
   \hat{\tilde{\kappa}} (x) &=& V^*\,(\hat{\kappa}\otimes\kappa)(x)\,V ,\eea
where $R$ and $V$ are given by
\bea (R\xi)(g,h) &=& (u_h\otimes\11)\,\xi(g,h) \\
   (V\xi)(g) &=& u_g\,\xi(g) .\eea
\eprop
\prf This is just the specialization of \cite[Th\'{e}or\`{e}me 2]{cann} to the
situation at hand. According to this theorem the von Neumann algebra underlying the 
dual of the crossed product Kac algebra $K\rtimes_\alpha G$ is 
$\hat{M}\otimes L^\infty(G)$ where $\hat{M}$ is the von Neumann algebra of $\hat{K}$.
In our case $M=L^\infty(G)$ such that $\hat{M}=\2L(G)$. The formulae for 
$\hat{\tilde{\Gamma}}$ and $\hat{\tilde{\kappa}}$ are stated in 
\cite[Proposition 4.10]{cann}. \qed\\
\rem If the group $G$ is not finite the quantum double is neither compact nor discrete,
for the weights $\tilde{\vp}, \hat{\tilde{\vp}}=\hat{\vp}\otimes\vp$ are both infinite.

We are now in a position to define a coaction of the dual double $\widehat{\2D(G)}$
on an algebra $A$, provided $A$ supports an action $\alpha$ and a coaction $\delta$
satisfying (\ref{coact}) (with $g$ replaced by $\lambda(g)$).
In order to remove the apparent asymmetry between $\alpha:A\times G\rightarrow A$ and
$\delta: A\rightarrow A\otimes\2L(G)$
we write the former as the homomorphism $\alpha: A\rightarrow A\otimes L^\infty(G)$
which maps $x\in A$ into $g\mapsto\alpha_g(x)\in L^\infty(G,A)$. We now show that the
maps $\alpha$ and $\delta$ can be put together to yield a coaction.
\bdefin The map 
$\Delta: A\rightarrow A\otimes\2L(G)\otimes L^\infty(G)=A\otimes\widehat{\2D(G)}$
is defined by
\be \Delta = (\imath_A\otimes\sigma)\circ(\alpha\otimes\imath_{\2L(G)})\circ\delta,\ee
where 
$\sigma: x\otimes y\mapsto y\otimes x$ is the flip map from $L^\infty(G)\otimes\2L(G)$
to $\2L(G)\otimes L^\infty(G)$.
\label{vn-coact}\edefin
\btheor The map $\Delta$ is a coaction of $\widehat{\2D(G)}$ on $A$, i.e.\ it 
satisfies
\be (\Delta\otimes\imath_{\hat{\2D}})\circ\Delta=
   (\imath_A\otimes\hat{\tilde{\Gamma}})\circ\Delta. \ee
\etheor
\prf Appealing to the isomorphism $A\otimes L^\infty(G)\cong L^\infty(G,A)$ we 
identify $A\otimes\2L(G)\otimes L^\infty(G)\otimes\2L(G)\otimes L^\infty(G)$ with
$L^\infty(G\times G,A\otimes\2L(G)\otimes\2L(G))$. We compute 
$(\Delta\otimes\imath)\circ\Delta(x)$ as follows
(abbreviating $\imath_{\2L(G)}$ by $\imath_\2L$)
\bea \lefteqn{((\Delta\otimes\imath)\circ\Delta(x))(g,h)=
   (\alpha_g\otimes\imath_\2L\otimes\imath_\2L)\circ(\delta\otimes\imath_\2L)\circ
   (\alpha_h\otimes\imath_\2L)\circ\delta(x)} \nn\\
 && = (\alpha_g\otimes\imath_\2L\otimes\imath_\2L)\circ(\alpha_h\otimes Ad\,\lambda_h
   \otimes\imath_\2L)\circ(\delta\otimes\imath_\2L)\circ\delta(x) \\
 && = (\imath_A\otimes Ad\,\lambda_h\otimes\imath_\2L)\circ
     (\alpha_{gh}\otimes\hat{\Gamma})\circ\delta(x) .\nn\eea
The second equality follows from the connection (\ref{coact}) between the action 
$\alpha$ and the coaction $\delta$ whereas the third derives from the defining 
property (\ref{coact-0}) of the coaction $\hat{\Gamma}$. Now
$(\alpha_{gh}\otimes\hat{\Gamma})\circ\delta(x)$ is seen to be nothing but
$[ (\11\otimes\sigma\otimes\11)\,(\hat{\Gamma}\otimes\Gamma)(x)\,(\11\otimes\sigma
\otimes\11)](g,h)$, and the adjoint action of $R$ in (\ref{gammahat}) is seen to
have the same effect as $Ad\,(\imath_A\otimes Ad\,\lambda_h\otimes\imath_\2L)$
due to $\rho(g)\in\2L(G)'$.\qed
\bprop The fixpoint algebra under the coaction $\Delta$, defined as
$A^\Delta = \{ x\in A \mid \Delta(x)=x\otimes\11_{\hat{\2D}} \}$, is given by
\be A^\Delta = A^\alpha \cap A^\delta ,\ee
where $A^\alpha,\;A^\delta$ are defined analogously.
\eprop
\prf Obvious consequence of Definition \ref{vn-coact}.\qed

The coaction of the dual double $\widehat{\2D(G)}$ on $A$ constructed above is exactly
the kind of output the theory of depth-2 inclusions \cite{lo2,en-ne} would give
when applied to the inclusion $A^{\2D(G)}\subset A$, which in the quantum field
theoretical application corresponds to $\2A(\2O)\subset\hat{\2F}(\2O)$. Nevertheless
it is perhaps not exactly what one might have desired from a generalization of the 
results of Section 4 to compact groups. At least to a physicist, some kind of bilinear
map $\gamma: A\times D(G)\rightarrow A$, as it was defined above for finite $G$, would
seem more intuitive. This map should be well defined on the whole algebra $A$.
Such a map can be constructed, provided the von Neumann double $\2D(G)$ is replaced by
its $C^*$-variant, which is uniquely defined by the above mentioned results 
\cite{ba-sk,en-va}. The details will be given in a subsequent publication. 

The representation theory of the quantum double in the (locally) compact case was
studied in a recent preprint \cite{ko-m} of which I became aware after completion of
the present work. An application of the results expounded there in analogy to Section
4 should be possible but is deferred for reasons of space.

\sectreset{Chiral Theories on the Circle}
For the foregoing analysis in this Chapter the split property for wedges was absolutely 
crucial. While this property has been proved only for free massive fields it is expected
to be true for all reasonable theories with a mass gap. For conformally invariant
theories in $1+1$ dimensions, however, it has no chance to hold. This is a consequence 
of the fact that two wedges $W_1\subset W_2$ `touch at infinity'. More precisely, there
is an element of the conformal group transforming $W_1,\, W_2$ into double cones
having a corner in common. For such regions there can be no interpolating type~$I$
factor, see e.g.\ \cite{bu}. On the other hand, for chiral theories on a circle,
into which a 1+1 dimensional conformal theory should factorize, an appropriate
kind of split property makes sense. For a general review of the framework, including
a proof of the split property from the finiteness of the trace of $e^{-\tau L_0}$,
we refer to \cite{fro3}. We restrict ourselves to a concise statement of the axioms.

For every interval $I$ on the circle such that $\ol{I}\ne S^1$, there is
a von Neumann algebra $\6A(I)$ on the common Hilbert space $\2H$. The 
assignment $I\mapsto\6A(I)$ fulfills isotony and locality:
\bea I_1\subset I_2 & \impl & \6A(I_1)\subset\6A(I_2) ,\\
   I_1\cap I_2=\emptyset & \impl & \6A(I_1)\subset\6A(I_2)' .\eea
Furthermore, there is a strongly continuous unitary representation of the
M\"obius group $SU(1,1)$ such that 
$\alpha_g(\6A(I))=Ad\,U(g)(\6A(I))=\6A(gI)$. Finally, the generator $L_0$ of the
rotations is supposed to be positive and the existence of a unique invariant vector
$\Omega$ is assumed.

Starting from these assumptions one can prove, among other important results, that
the local algebras $\6A(I)$ are factors of type $III_1$ for which the vacuum is
cyclic and separating. Furthermore, Haag duality \cite{fro3} is fulfilled 
automatically:
\be \6A(I)'=\6A(I') .\ee
Given a chiral theory in its defining (vacuum) representation $\pi_0$ one may 
consider inequivalent representations. An important first result \cite{bmt} states 
that all positive energy representations are locally equivalent to the vacuum 
representation, i.e.\ $\pi\restr\6A(I)\cong{\pi_0}\restr\6A(I)\ \forall I$.
This implies that all superselection sectors are of the DHR type and can be
analyzed accordingly \cite{fre1,frs2}. As a means of studying the superselection
theory of a model it has been proposed \cite{schroer} to examine the inclusion
\be \6A(I_1)\vee\6A(I_3)\subset(\6A(I_2)\vee\6A(I_4))'=
   \6A(I_{341})\wedge\6A(I_{123}) ,\label{1234}\ee
where $I_{1,\ldots, 4}$ are quadrants of the circle and 
$I_{ijk}=\ol{I_i\cup I_j\cup I_k}$:
\be\ba{c}\begin{picture}(100,70)(-50,-35)\thicklines
\put(0,0){\circle{50}}
\put(-3,30){$I_1$}
\put(-3,-40){$I_3$}
\put(-40,-3){$I_2$}
\put(30,-3){$I_4$}
\put(-10,-10){\line(-1,-1){10}}
\put(-10,10){\line(-1,1){10}}
\put(10,-10){\line(1,-1){10}}
\put(10,10){\line(1,1){10}}
\end{picture}\ea\ee
At least for strongly additive theories, where $\6A(I_1)\vee\6A(I_2)=\6A(I)$
if $\ol{I_1\cup I_2}=I$, the inclusion (\ref{1234}) is easily seen to be
irreducible. In the presence of nontrivial superselection sectors this inclusion
is strict as the intertwiners between endomorphisms localized in $I_1, I_3$,
respectively, are contained in the larger algebra of (\ref{1234}) by Haag duality
but not in the smaller one. Furthermore, for rational theories the inclusion
(\ref{1234}) is expected to have finite index.

While we have nothing to add in the way of model independent analysis the techniques
developed in the preceding sections can be applied to a large class of interesting
models. These are chiral nets obtained as fixpoints of a larger one under the
action of a group. I.~e. we start with a net $I\mapsto\2F(I)$ on the Hilbert
space $\2H$ fulfilling isotony and locality, the latter possibly twisted.
The M\"obius group $SU(1,1)$ and the group $G$ of inner symmetries are unitarily
represented with common invariant vector $\Omega$. Again, the net $\2F$ is 
supposed to fulfill the split property (with the obvious modifications due to the
different geometry). The net $I\mapsto\6A(I)$ is now defined by
$\2A(I)=\2F(I)\wedge U(G)'$ and $\6A(I)=\2A(I)\restr\2H_0$ where 
$\2H_0$ is the space of G-invariant vectors. The proof of Haag duality for
chiral theories referred to above applies also to the net $\6A$, implying that
there is no analogue of the violation of duality for the fixpoint net as occurs in 1+1
dimensions. This is easily understood as a consequence of the fact that the spacelike
complement of an interval is again an interval, thus connected. However, our
methods can be used to study the inclusion (\ref{1234}).

It is clear that due to the split property
\be \6A(I_1)\vee\6A(I_3)\cong {\2F(I_1)\otimes\2F(I_3)}^{G\times G}
   \restr\2H_0\otimes\2H_0 .\ee
Our aim will now be to compute $(\6A(I_2)\vee\6A(I_4))'$.
In analogy to the 1+1 dimensional case we use the split property to construct
unitaries $Y_1,\ldots, Y_4:\,\2H\to\2H\otimes\2H$ implementing the following
isomorphisms:
\be Y_i \, F_i F^t_{i+2} \, Y_i^* = F_i \otimes F^t_{i+2} \ \
  \forall F_i\in\2F(I_i) .\ee
(One easily checks that $Y_{i+2}=T\,Y_i$ where $T\,x\otimes y=y\otimes x$.)
These unitaries can in turn be used to define local implementers of the gauge
transformations 
\be U_i(g)=Y_i^*\, (U(g)\otimes\11)\, Y_i \ee
with the localization $U_i(g)\in\2F(I_{i+2})'$. (The index arithmetic takes
place modulo $4$.) These operators satisfy
\bea Ad\,U_i(g)\restr\2F(I_i) &=& \alpha_g ,\\
  \left[ U_i(g),U_{i+2}(h) \right] &=& 0 ,\\
  U_i(g)\,U_{i+2}(g) &=& U(g) .\eea
In a manner analogous to the proof of Lemma \ref{lem1} one shows ($\2F_i\equiv\2F(I_i)$
etc.)
\be (\2A_2\vee\2A_4)'=(\2F_2\vee\2F_4)'\vee U_2(G)''\vee U_4(G)'' .\ee

At this point we strengthen the property of Haag duality for the net $\2F$ by
requiring
\be (\2F_1\vee\2F_3)'=(\2F_2\vee\2F_4)^t ,\label{nosect}\ee
which by the above considerations amounts to $\2F$ having no nontrivial
superselection sectors. This condition is fulfilled, e.g., by the CAR algebra on
the circle which also possesses the split property. The chiral Ising model as 
discussed in \cite{boc} is covered by our general framework (with the group 
$\7Z_2$).

While (\ref{nosect}) is a strong restriction it is the same as in
\cite{dvvv} where the larger theory was supposed to be `holomorphic'. At this
place it might be appropriate to emphasize that the requirement of (twisted) Haag
duality (\ref{twduality}) made above when considering 1+1 dimensional theories
by no means excludes nontrivial superselection sectors.

Making use of (\ref{nosect}) we can now state quite explicitly how 
$(\6A_2\vee\6A_4)'$ looks. In analogy to
Theorem \ref{thm1} we obtain
\be (\6A_2\vee\6A_4)'=m\left(\2F_1\vee\2F_3\vee U_2(G)''\right)\restr\2H_0 .\ee
Again, using (\ref{nosect}) one can check that $\alpha_2(g)=Ad\,U_2(g)$ restrict to
automorphisms of $\2F_1\vee\2F_3$ rendering the algebra 
$\2F_1\vee\2F_3\vee U_2(G)''$ a crossed product. Recalling
\be \6A_1\vee\6A_3=m(\2F_1)\vee m(\2F_3)\restr\2H_0 \ee
we have the following natural sequence of inclusions:
\be \6A_1\vee\6A_3 \subset m(\2F_1\vee\2F_3)\restr\2H_0 \subset 
   (\6A_2\vee\6A_4)' \label{lowdim}\ee
both of which have index $|G|$. It is interesting to remark that the  intermediate
algebra $m(\2F_1\vee\2F_3)\restr\2H_0$ equals
$(m(\2F_2\vee\2F_4)\restr\2H_0)'$. For general chiral theories the existence 
of such an intermediate subfactor between $\6A_1\vee\6A_3$ and $(\6A_2\vee\6A_4)'$ 
is not known. 
In the case of $G$ being abelian where the $U_i(g)$ are invariant
under global gauge transformations we obtain a square structure similar to the
one encountered in Section 3.
\be\ba{ccc} \6A_1\vee\6A_3\vee U_2(G)'' & \subset & (\6A_2\vee\6A_4)' \\ 
   \cup & & \cup \\
   \6A_1\vee\6A_3 & \subset & m(\2F_1\vee\2F_3)\restr\2H_0
.\ea\label{square2}\ee
%The observation that $\alpha_1(g)=Ad\,U_1(g)$ acts automorphically on 
%$\2F_1\vee\2F_3$, too, with
%$\6A_1\vee\6A_3=(m(\2F_1\vee\2F_3)\restr\2H_0)\wedge U_1(G)'$
%shows that the inclusion 

It may be instructive to compare the above result with the situation prevailing in
$2+1$ or more dimensions. There, as already mentioned in the introduction, the 
superselection theory for localized charges is isomorphic to the representation 
theory of a (unique) compact group. Furthermore, there is a net of field algebras 
acted upon by this group, such that the observables arise as the fixpoints. 
The analogue of the inclusion (\ref{1234}) then is
\be \6A(\2O_1)\vee\6A(\2O_2)\subset\6A(\2O_1'\cap\2O_2')' ,\label{highdim}\ee
where $\2O_1,\,\2O_2$ are spacelike separated double cones. Under natural
assumptions it can be shown that the larger algebra equals 
$m(\2F(\2O_1)\vee\2F(\2O_2))\restr\2H_0$, implying that the inclusion
(\ref{highdim}) is of the type 
$(\2F_1\otimes\2F_2)^{G\times G}\subset(\2F_1\otimes\2F_2)^{\mbox{Diag}(G)}$
just as the first one in (\ref{lowdim}). That the index of the inclusion (\ref{1234})
is $|G|^2$ instead of $|G|$ as for (\ref{highdim}) is a consequence of the low
dimensional topology comparable to the phenomena occurring in $1+1$ dimensions.

\vspace{1cm}
\noindent{\it Note added in proof.}
In Appendix C we claimed that the combination of Haag duality and the
split property for wedges is weaker than the requirement of absence of charged sectors
which was made in \cite{dvvv} where conformal orbifold theories were considered.
After submission of this paper we discovered that this claim is wrong! While this does
not affect any result of the present work it shows that the analysis of massive models 
based on the former assumptions is even stronger related to the one in \cite{dvvv} than
expected. Furthermore, if the vacuum sector satisfies HD+SPW then Haag duality holds in
all irreducible locally normal representations. In particular, on can replace
`simple sector' by `irreducible $\2A$-stable subspace of $\2H$' in Theorem 3.10.
The proofs as well as applications to the theory of quantum solitons will be found in
\cite{mue3}. 


\begin{thebibliography}{99}
\bibitem{adler} Adler, C.: Braid group statistics in two-dimensional \qft,
   \rmp {\bf 8}(1996)907
\bibitem{ac} Altschuler, D., Coste, A.: Invariants of 3-manifolds from finite groups,
   {\it in:} Proc. XXth Conf. Diff. Geom. Meth. Theor. Phys., New York, 1991
\bibitem{araki} Araki, H. : A lattice of von Neumann algebras associated with the 
  quantum theory of a free Bose field, J. Math. Phys. {\bf 4}(1963)1343
\bibitem{araki2} Araki, H.: On the XY-model on two-sided infinite chain, Publ. RIMS
   {\bf 20}(1984)277
\bibitem{ba-sk} Baaj, P. S., Skandalis, G.: Unitaires multiplicatifs et dualit\'{e}
   pour les produits crois\'{e}s de $C^*$-alg\`{e}bres,
   Ann. Sci. \'{E}NS {\bf 26}(1993)425
\bibitem{b-wp} Bais, F. A., de Wild Propitius, M.: Quantum groups in the Higgs phase,
   Theor. Math. Phys. {\bf 98}(1994)357
\bibitem{bern} Bernard, D., LeClair, A.: The quantum double in integrable \qft, 
   \npb {\bf 399}(1993)709
\bibitem{bisch} Bisch, D., Haagerup, U.: Composition of subfactors: New examples of 
   infinite depth subfactors, Ann. Sci. \'{E}NS {\bf 29}(1996)329
\bibitem{boc} B\"ockenhauer, J.: Localized endomorphisms of the chiral Ising model,
   \cmp {\bf 177}(1996)265
\bibitem{bonn2} Bonneau, P.: Topological quantum double, \rmp {\bf 6}(1994)305
\bibitem{bu} Buchholz, D.: Product states for local algebras, \cmp {\bf 36}(1974)287
\bibitem{buwi} Buchholz, D., Wichmann, E. H.: Causal independence and the energy-level
   density in local \qft, \cmp {\bf 106}(1986)321
\bibitem{bmt} Buchholz, D., Mack, G., Todorov, I.: The current algebra on the circle
   as a germ of local field theories, \npb (Proc. Suppl.){\bf 5B}(1988)20
\bibitem{bdl} Buchholz, D., Doplicher, S., Longo, R.: On Noether's theorem in \qft,
   Ann. Phys. {\bf 170}(1986)1
\bibitem{bdlr} Buchholz, D., Doplicher, S., Longo, R., Roberts, J. E.: A new look at
   Goldstone's theorem, \rmp Special Issue(1992)49
\bibitem{bu-p} Buchholz, D.: private communication
\bibitem{cann} De Canni\`{e}re, J.: Produit crois\'{e} d'une alg\`{e}bre de Kac
   par un groupe localement compact, Bull. Soc. Math. France {\bf 107}(1979)337
\bibitem{cohen} Cohen, M., Fishman, D.: Hopf algebra actions, J. Algebra 
   {\bf 100}(1986)363
\bibitem{dvvv} Dijkgraaf, R., Vafa, C., Verlinde, E., Verlinde, H.: The operator
   algebra of orbifold models, \cmp {\bf 123}(1989)485
\bibitem{dpr} Dijkgraaf, R., Pasquier, V., Roche, P.: Quasi Hopf algebras, group
   cohomology and orbifold models, \npb (Proc. Suppl.){\bf 18B}(1990)60
\bibitem{dhr1} Doplicher, S., Haag, R., Roberts, J. E.: Fields, observables and
   gauge transformations I, \cmp {\bf 13}(1969)1
\bibitem{dhr2} Doplicher, S., Haag, R., Roberts, J. E.: Fields, observables and
   gauge transformations II, \cmp {\bf 15}(1969)173
\bibitem{dhr3} Doplicher, S., Haag, R., Roberts, J. E.: Local observables and particle
   statistics I, \cmp {\bf 23}(1971)199
\bibitem{dhr4} Doplicher, S., Haag, R., Roberts, J. E.: Local observables and particle
   statistics II, \cmp {\bf 35}(1974)49
\bibitem{dr1} Doplicher, S., Roberts, J. E.: Fields, statistics and non-abelian gauge
   groups, \cmp {\bf 28}(1972)331
\bibitem{dopl} Doplicher, S.: Local aspects of superselection rules, 
   \cmp {\bf 85}(1982)73
\bibitem{dl} Doplicher, S., Longo, R.: Standard and split inclusions of von
   Neumann algebras, Invent. Math. {\bf 75}(1984)493
\bibitem{dr2} Doplicher, S., Roberts, J. E.: Why there is a field algebra with a 
   compact gauge group describing the superselection structure in particle physics, 
   \cmp {\bf 131}(1990)51
\bibitem{dri} Driessler, W.: On the type of local algebras in \qft, 
   \cmp {\bf 53}(1977)295
\bibitem{drin1} Drinfel'd, V. G.: Quantum groups, {\it in:} Proc. Int. Congr. Math,
   Berkeley 1986
\bibitem{drin2} Drinfel'd, V. G.: Quasi-Hopf algebras, Leningrad Math.J. 
   1(1989)1419-1457
\bibitem{en-schw} Enock, M., Schwartz, J.-M.: Kac algebras and duality of locally
   compact groups, Springer, 1992
\bibitem{en-va} Enock, M., Vallin, J.-M.: $C^*$-alg\`{e}bres de Kac et alg\`{e}bres 
   de Kac, Proc. Lond. Math. Soc. {\bf 66}(1993)619
\bibitem{en-ne} Enock, M., Nest, R.: Irreducible inclusions of factors, multiplicative
   unitaries, and Kac algebras, \jfa {\bf 137}(1996)466
\bibitem{frs1} Fredenhagen, K., Rehren, K.-H., Schroer, B.: Superselection sectors 
   with braid group statistics and exchange algebras I. General theory, \cmp
   {\bf 125}(1989)201
\bibitem{frs2} Fredenhagen, K., Rehren, K.-H., Schroer, B.: Superselection sectors 
   with braid group statistics and exchange algebras II. Geometric aspects and 
   conformal covariance, \rmp Special Issue(1992)113
\bibitem{fre1} Fredenhagen, K.: Generalizations of the theory of superselection
   sectors, {\it in}: \cite{K}
\bibitem{fre2} Fredenhagen, K.: Superselection sectors in low dimensional \qft,
   J. Geom. Phys. {\bf 11}(1993)337
\bibitem{fro1} Fr\"ohlich, J.: New super-selection sectors (`Soliton-states') in 
   two-dimensional Bose quantum field models, \cmp {\bf 47}(1976)269
\bibitem{fro2} Fr\"ohlich, J.: Statistics of fields, the Yang-Baxter equation, and the
   theory of knots and links, {\it in:} t'Hooft, G. et al.\ (eds.): Nonperturbative
   quantum field theory, Carg\`{e}se Summer School 1987
\bibitem{fuchs} Fuchs, J.: Fusion rules in conformal field theory, 
   Fortschr. Phys. {\bf 42}(1994)1
\bibitem{fro3} Gabbiani, F., Fr\"ohlich, J.: Operator algebras and conformal field 
   theory, \cmp {\bf 155}(1993)569
\bibitem{str} Gallone, F., Sparzani, A., Streater, R. F., Ubertone, C.:
   Twisted condensates of quantized fields, J. Phys. {\bf A19}(1986)241
\bibitem{gs} G\'{o}mez, C., Sierra, G.: Quantum group meaning of the Coulomb gas,
   Phys. Lett. {\bf B240}(1990)149
\bibitem{haag1} Haag, R., Kastler, D.: An algebraic approach to \qft, J. Math. Phys.
   {\bf 3}(1964)848
\bibitem{haag2} Haag, R.: Local Quantum Physics, Springer Texts and Monographs
   in Physics, 1992
\bibitem{haage} Haagerup, U.: On the dual weights for crossed products of von Neumann
   algebras II. Application of operator valued weights, Math. Scand. {\bf 43}(1978)119
\bibitem{haga} Haga, Y.: Crossed products of von Neumann algebras by compact groups,
   T\^{o}hoku. Math. J. {\bf 28}(1976)511
\bibitem{K} Kastler, D. (ed.): The algebraic theory of superselection sectors. 
   Introduction and recent results, World Scientific, 1990
\bibitem{km} K\"oberle, R., Marino, E. C.: Duality, mass spectrum and vacuum
   expectation values, Phys. Lett. {\bf B126}(1983)475
\bibitem{ko-m} Koornwinder, T. H., Muller, N. M.: The quantum double of a (locally)
   compact group, q-alg/9605044
\bibitem{landst} Landstad, M. B., Phillips, J., Raeburn, I., Sutherland, C. E.:
   Representations of crossed products by coactions and principal bundles,
   Trans. AMS {\bf 299}(1987)747
\bibitem{lehm} Lehmann, H., Stehr, J.: The Bose field structure associated with a
   free massive Dirac field in one space dimension, DESY-preprint 76-29
\bibitem{lo1} Longo, R.: Index of subfactors and statistics of quantum fields I,
   \cmp {\bf 126}(1989)217
\bibitem{lo2} Longo, R.: A duality for Hopf algebras and for subfactors I, \cmp 
   {\bf 159}(1994)133
\bibitem{mue3} M\"uger, M.: Superselection structure of massive \qfts\ in $1+1$
   dimensions, hep-th/9705019
\bibitem{mue5} M\"uger, M.: A strong version of causal independence in two 
   dimensional quantum field theory, in preparation
\bibitem{khr} Rehren, K.-H.: Braid group statistics and their superselection rules,
   {\it in}: \cite{K}
\bibitem{reshet} Reshetikhin, N. Yu., Semenov-Tian-Shansky, M. A.: Quantum R-matrices
   and factorization problems, J. Geom. Phys. {\bf 5}(1988)533
\bibitem{rt} Reshetikhin, N. Yu., Turaev, V. G.: Invariants of 3-manifolds via link
   polynomials and quantum groups, Invent. Math. {\bf 103}(1991)547; Turaev, V. G.: 
   Modular categories and 3-manifold invariants, Int. Journ. Mod. Phys. 
   {\bf B6}(1992)1807
\bibitem{rob} Roberts, J. E.: Spontaneously broken gauge symmetries and superselection
   rules, {\it in:} Gallavotti, G. (ed.): Proc. International School of Mathematical 
   Physics, Camerino 1974   
\bibitem{rob2} Roberts, J. E.: Cross products of von Neumann algebras by group
   duals, Sympos. Math. {\bf 20}(1976)335
\bibitem{schl} Schlingemann, D.: On the existence of kink-(soliton-) states in \qft,
   \rmp {\bf 8}(1996)1187
\bibitem{schroer} Schroer, B.: Recent developments of algebraic methods in \qfts,
   Int. Journ. Mod. Phys. {\bf B6}(1992)2041
\bibitem{stra} Str\v{a}til\v{a}, S.: Modular theory in operator algebras,
   Abacus Press, 1981
\bibitem{swi} Swieca, J. A.: Fields with generalized statistics: An exercise in order
   and disorder in two dimensional systems, {\it in:} Turko, L. et al.\ (eds.): 
   Fundamental Interactions, Karpacz Winter School 1980
\bibitem{szlvec} Szlach\'{a}nyi, K., Vecserny\'{e}s, P.: Quantum symmetry and braid
   group statistics in G-spin models, \cmp {\bf 156}(1993)127
\bibitem{szym1} Szyma\'{n}ski, W., Peligrad, C.: Saturated actions of finite 
   dimensional Hopf algebras on $C^*$-algebras, Math. Scand. {\bf 75}(1994)217
\bibitem{szym2} Szyma\'{n}ski, W.: Finite index subfactors and Hopf algebra crossed
   products, Proc. Amer. Math. Soc. {\bf 120}(1994)519
\bibitem{takes} Takesaki, M.: Duality for crossed products and the structure of von
   Neumann algebras of type $III$, Acta Math. {\bf 131}(1973)249
\bibitem{v} Verlinde, E.: Fusion rules and modular transformations in 2D conformal 
   field theory, \npb {\bf 300}(1988)360
\end{thebibliography}
\end{document}